\documentclass[12pt]{article}

\usepackage{scicite}
\usepackage{graphicx}

\usepackage{times}

\newcommand{\ltsim}{\raisebox{-.5ex}{$\;\stackrel{<}{\sim}\;$}}
\newcommand{\gtsim}{\raisebox{-.5ex}{$\;\stackrel{>}{\sim}\;$}}

\newcommand{\Msun}{M_{\odot}}
\newcommand{\Msol}{\Msun}
\newcommand{\mpyr}{\Msun\,{\rm yr}^{-1}}
\newcommand{\ergs}{{\rm erg\,s}^{-1}}
\newcommand{\kev}{keV}
\newcommand{\kms}{{\rm km\,s}^{-1}}
\newcommand{\NDunit}{{\rm Mpc}^{-3}}
\def\arcsec{\hbox{$^{\prime\prime}$}}
\newcommand{\mic}{$\mu$m}
\newcommand{\cmii}{{\rm cm}^{-2}}

\newcommand{  \Lya      }{Ly$\alpha$}
\newcommand{  \Hbeta    }{H$\beta$}
\newcommand{  \hb    	}{{\rm H}\beta}
\newcommand{\oiii	}{[O\,\textsc{iii}]}

\newcommand{  \HeIIop   }{He\,\textsc{ii}\,$\lambda4686$}
\newcommand{  \feii     }{Fe\,\textsc{ii}}
\newcommand{  \mgii     }{{\rm Mg}\,\textsc{ii}}
\newcommand{  \MgII     }{Mg\,\textsc{ii}\,$\lambda2798$}
\newcommand{  \CIII     }{C\,\textsc{iii}]\,$\lambda1909$}
\newcommand{  \civ      }{{\rm C}\,\textsc{iv}}
\newcommand{  \CIV      }{C\,\textsc{iv}\,$\lambda1549$}
\newcommand{  \Siliv    }{{\rm Si}\,\textsc{iv}}
\newcommand{  \SilIVuv  }{Si\,\textsc{iv}\,$\lambda1400$}
\newcommand{  \AlIII   }{Al\,\textsc{iii}\,$\lambda1857$}

\newcommand{  \lamLlam  }{\lambda L_{\lambda}}
\newcommand{  \Lop      }{L_{5100}}
\newcommand{  \Lbol     }{L_{\rm bol}}
\newcommand{  \Lagn     }{L_{\rm AGN}}
\newcommand{  \mbh      }{M_{\rm BH}}
\newcommand{  \lledd    }{L/L_{\rm Edd}}

\newcommand{ \fwhm  }{{\rm FWHM}} 
\newcommand{ \fwhb  }{{\rm FWHM}\left(\hb\right)}
\newcommand{ \fwmg  }{{\rm FWHM}\left(\mgii\right)}
\newcommand{\sigBLR }{\sigma_{\mbox{\tiny BLR}}}
\newcommand{\fbolopt}{f_{\rm bol}\left(5100{\rm \AA}\right)}
\newcommand {\RBLR  }{R_{\rm BLR}}

\newcommand{  \mstar    }{M_{*}} 
\newcommand{  \mgal     }{M_{*}}

\newcommand{  \mmsmall  }{M_{\rm BH}/M_{*}}

\newcommand{  \mseed    }{M_{\rm seed}}

\newcommand{\kband}{\textit{K}-band}
\newcommand{  \spitzer }  {{\it Spitzer}}
\newcommand{  \herschel} {{\it Herschel}}
\newcommand{  \chandra }  {{\it Chandra}}
\newcommand{  \xmm     }  {{\it XMM-Newton}}

\newcommand{\mysobj}{CID--947}

\newenvironment{sciabstract}{%
\begin{quote} \bf}
{\end{quote}}

\newcounter{lastnote}

\title{An Over-Massive Black Hole in a Typical Star-Forming Galaxy, 2 Billion Years After the Big Bang}

\author
{Benny Trakhtenbrot,$^{1\ast}$ 
C. Megan Urry,$^{2,3,4}$ 
Francesca Civano,$^{3,5}$ \\
David J. Rosario,$^{6}$ 
Martin Elvis,$^{5}$ 
Kevin Schawinski,$^{1}$\\
Hyewon Suh,$^{5,7}$
Angela Bongiorno,$^{8}$
Brooke D. Simmons$^{9}$
\\
\\
\normalsize{$^{1}$Department of Physics, Institute for Astronomy, ETH Zurich,}\\
\normalsize{Wolfgang-Pauli-Strasse 27, Zurich 8093, Switzerland}\\
\normalsize{$^{2}$Department of Physics, Yale University, PO Box 208120, New Haven, CT 06520-8120, USA}\\
\normalsize{$^{3}$Yale Center for Astronomy and Astrophysics, 260 Whitney ave., New Haven, CT 06520-8121, USA}\\
\normalsize{$^{4}$Department of Astronomy, Yale University, PO Box 208101, New Haven, CT 06520-8101, USA}\\
\normalsize{$^{5}$Harvard-Smithsonian Center for Astrophysics, 60 Garden st., Cambridge, MA 02138, USA}\\
\normalsize{$^{6}$Max-Planck-Institut f\"{u}r Extraterrestrische Physik (MPE), Postfach 1312, 85741 Garching, Germany}\\
\normalsize{$^{7}$Institute for Astronomy, University of Hawaii, 2680 Woodlawn Drive, Honolulu, HI 96822, USA}\\
\normalsize{$^{8}$INAF-Osservatorio Astronomico di Roma, Via di Frascati 33, I-00040 Monteporzio Catone, Rome, Italy}\\
\normalsize{$^{9}$Oxford Astrophysics, Denys Wilkinson Building, Keble Road, Oxford OX1 3RH, UK}\\
\\
\normalsize{$^\ast$Corresponding author. E-mail: benny.trakhtenbrot@phys.ethz.ch}
}

\date{}

\begin{document} 


\baselineskip24pt


\maketitle

\paragraph*{}
\label{sec:abstract}


\begin{sciabstract}

Supermassive black holes (SMBHs) and their host galaxies are generally thought
to coevolve, so that the SMBH achieves up to about 0.2 to 0.5\% of the host galaxy
mass in the present day. 
The radiation emitted from the growing SMBH is expected to affect star formation throughout the host galaxy. 
The relevance of this scenario at early cosmic epochs is not yet established.
We present spectroscopic observations of a galaxy at redshift $z = 3.328$, which hosts an actively accreting, extremely massive BH, in its final stages of growth.
The SMBH mass is roughly one-tenth the mass of the entire host galaxy, suggesting that it has grown much more efficiently than the host, contrary to models of synchronized coevolution. 
The host galaxy is forming stars at an intense rate, despite the presence of a SMBH-driven gas outflow.

%
%
\end{sciabstract}



Several lines of observational evidence, spanning a wide range of cosmic epochs, have led to a commonly accepted picture wherein supermassive black holes (SMBHs, $\mbh>10^6\,\Msol$; $\Msol$ is the solar mass) coevolve with their host galaxies \cite{Ferrarese2000,Gebhardt2000_Msig,Zheng2009_BHAD_SFRD,KormendyHo2013_MM_Rev}.
Moreover, energy- and/or momentum-driven ``feedback'' from accreting SMBHs (Active Galactic Nuclei; AGN) is thought to quench star formation in the host galaxy \cite{Fabian2012_feedback_rev}. 
To directly test the relevance of such scenarios at early cosmic epochs (high redshifts, $z$) requires the most basic properties of SMBHs and their hosts, including masses and growth rates, to be observed.
Several observational studies found that at $z\ltsim2$ (more than 3.3 billion years after the Big Bang), the typical BH-to-stellar mass ratio, $\mmsmall$, increases towards higher redshifts \cite{Merloni2010,Decarli2010_MM_evo,Bennert2011}, suggesting that some SMBHs were able to gather mass more efficiently, or faster, than the stellar populations in their hosts.
To date, measurements of $\mbh$ at earlier epochs ($z>2$) have only been conducted for small samples of extremely luminous objects [$\Lagn > 10^{46}\,\ergs$ \cite{Shemmer2004,Netzer2007_MBH,DeRosa2011,Trakhtenbrot2011}] representing a rare subset of all accreting SMBHs, with number densities of order $1$ to $10$ per Gpc$^{3}$ [i.e., $\sim10^{-9}$ to $10^{-8}\,\NDunit$ \cite{Masters2012}].
Moreover, the high AGN luminosities in such sources overwhelm the host galaxy emission and prohibit a reliable determination of $\mstar$, and therefore of $\mmsmall$.
We initiated an observational campaign aimed at estimating $\mbh$ in x-ray--selected, unobscured $z\sim3$ to $4$ AGN within the Cosmic Evolution Survey field [COSMOS; \cite{Scoville2007_COSMOS_overview}].
Such sources have lower AGN luminosities and are more abundant than the aforementioned luminous sources by factors of $100$ to $1000$ \cite{Civano2011_hiz,Masters2012} and thus form a more representative subset of the general AGN population. 
Moreover, the fainter AGN luminosities and rich multiwavelength coverage of AGN within the COSMOS field enable reliable measurements of the mass and growth rate of the stellar populations in the host galaxies ($\mgal$ and star-formation rate, SFR).

\paragraph*{}
\label{sec:results}

\begin{figure*}[t]
\centering
\includegraphics[width=0.85\textwidth]{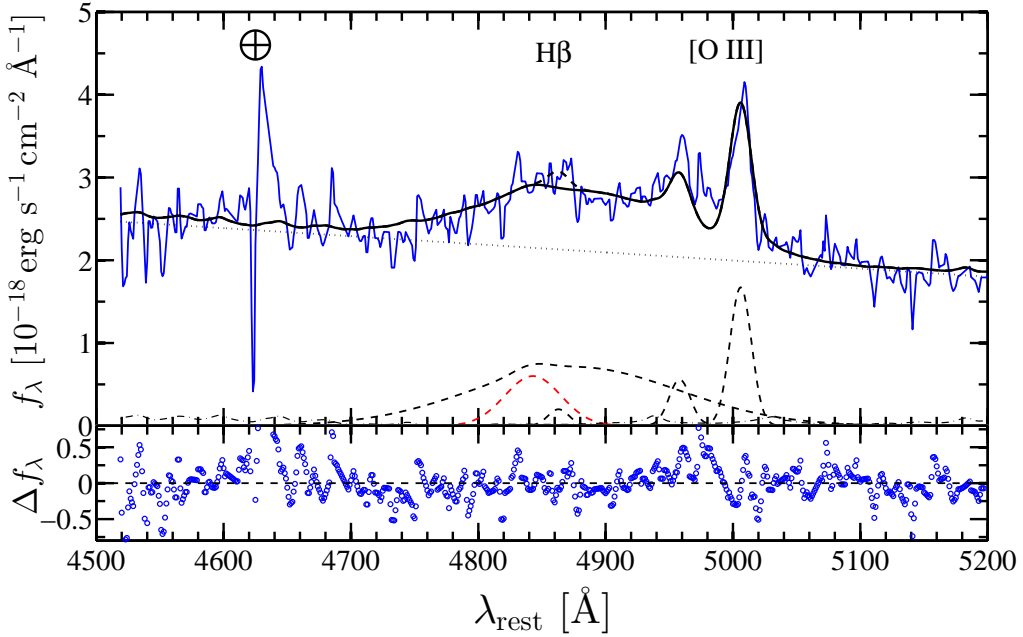} 
\caption{
{\bf The observed Keck/MOSFIRE spectrum and best-fit model for the \Hbeta\ emission complex of \mysobj.}
The data are modeled with a linear continuum (dotted), a broadened iron template (dot-dashed) and a combination of broad and narrow Gaussians (dashed), which correspond to the \Hbeta\ and \oiii\ emission lines 
(see section \ref{SM_sec_analysis} in the supplementary materials for details regarding the spectral modeling).
The broad component of \Hbeta\ has a full width at half maximum of $\fwhb=11330\,\kms$, which results in $\mbh=6.9\times10^9\,\Msol$ and $\mmsmall=1/8$.
The red dashed line illustrates an alternative scenario, in which the SMBH mass derived from the \Hbeta\ line width would result in $\mmsmall=1/100$ [i.e., ${\rm FWHM}\left(\hb\right)=3218\,\kms$], clearly at odds with the data.
The spike at $\lambda_{\rm rest}\simeq4640$ \AA\ is due to a sky feature.
The bottom panel shows the residuals of the best fit model.
}
\label{fig:CID_947_spec_fit_resid}
\end{figure*}

\mysobj\ is an x-ray--selected, unobscured AGN at  $z=3.328$, detected in both \xmm\ and \chandra\ x-ray imaging data of the COSMOS field [see Fig.~\ref{fig:CID_947_zCOSMOS} and sections \ref{SM_sec_SED} and \ref{SM_sec_BAL} in the supplementary materials \cite{SM_note}].  
We obtained a near-infrared (IR) \kband\ spectrum of \mysobj\ using the MOSFIRE instrument at the W. M. Keck telescope, 
which at $z=3.328$ covers the hydrogen \Hbeta\ broad emission line (see details in section \ref{SM_sec_Kband_spec} in the supplementary materials).
The calibrated spectrum shows a very broad \Hbeta\ emission line, among other features (Fig.~\ref{fig:CID_947_spec_fit_resid}). 
Our spectral analysis indicates that the monochromatic AGN luminosity at rest-frame 5100 \AA\ is $\Lop=3.58^{+0.07}_{-0.08}\times10^{45}\,\ergs$.
The typical line-of-sight velocity, i.e.\ the full-width at half-maximum of the line, is $11330^{+800}_{-870}\,\kms$ (see section \ref{SM_sec_analysis} in the supplementary materials). 
By combining this line width with the observed $\Lop$ and relying on an empirically calibrated estimator for $\mbh$, based on the virial motion of ionized gas near the SMBH \cite{Shen2013_rev}, we obtain $\mbh=6.9^{+0.8}_{-1.2}\times10^{9}\,\Msun$.
All the reported measurement-related uncertainties are derived by a series of simulations and represent the 16th and 84th quantiles of the resulting distributions.
These simulations indicate a SMBH mass larger than $3.6\times10^9\,\Msun$ at the 99\% confidence level (see sections \ref{SM_sec_analysis} and \ref{SM_sec_M_LLedd} for more details).
Determinations of $\mbh$ from single-epoch spectra of the \Hbeta\ emission line are known to also be affected by significant systematic uncertainties, of up to $\sim0.3$ to $0.4$ dex.
For a detailed discussion of some of the systematics and related issues, see \S\ref{SM_sec_M_LLedd} in the supplementary materials.
This high $\mbh$ is comparable with some of the most massive BHs known to date in the local universe \cite{McConnell2012_BCGs},or with the masses
of the biggest BHs in the much rarer, more luminous AGN at $z\sim2$ to $4$ [e.g., \cite{Shemmer2004}].
The bolometric luminosity of \mysobj\ is in the range $\Lbol\simeq\left(1.1-2.2\right)\times10^{46}\,\ergs$, estimated either from the observed optical luminosity or the multiwavelength spectral energy distribution.
Combined with the measured $\mbh$, we derive a normalized accretion rate of $\lledd\simeq0.01$ to $0.02$. 
This value is lower, by at least an order of magnitude, than the accretion rates of known SMBHs at $z\sim3.5$ [e.g., \cite{Shemmer2004,Netzer2007_MBH}].
Further assuming a standard radiative efficiency of 10\%, we obtain an $e$-folding time scale for the SMBH mass of at least $2.1\times 10^9$ (Gy; see section \ref{SM_sec_M_LLedd}), which is longer than the age of the universe at $z=3.328$.
By contrast, even the most extreme models for the emergence of ``seed'' BHs predict masses no larger than $\mseed\sim 10^{6}\,\Msol$ at $z\sim10$ to $20$ [e.g., \cite{Volonteri2010_rev}].
Therefore, the SMBH powering \mysobj\ had to grow at much higher accretion rates and at a high duty cycle in the past, to account for the high observed $\mbh$ only 1.7 Gyr after $z\simeq20$.
\mysobj\ could have evolved from a parent population similar to the fast-growing SMBHs observed in $z\gtsim5$ quasars, which have $\lledd\sim0.5$ to $1$ and $\mbh\simeq10^9\,\Msol$ [e.g., \cite{Trakhtenbrot2011,DeRosa2011}].
The requirement for a high accretion rate in the very recent past is supported by the clear presence of a high-velocity outflow of ionized gas, observed in the rest-frame ultraviolet spectrum of the source (fig.~\ref{fig:CID_947_zCOSMOS}). 
The broad absorption features of \CIV\ and \SilIVuv\ have maximal velocities of $v_{\rm max}\simeq12,000\,\kms$. 
Assuming that this outflow is driven by radiation pressure, these velocities require accretion rates of $\lledd\gtsim0.1$, as recently as $10^5$ to $10^6$ years before the observed epoch (see section \ref{SM_sec_BAL}).
We conclude that the SMBH powering \mysobj\ is in the final stages of growth and that we are witnessing the shut-down of accretion onto one of the most massive BHs known to date.

The rich collection of ancillary COSMOS multiwavelength data available for \mysobj\ enables us to study the basic properties of its host galaxy (see details in section \ref{SM_sec_SED} in the supplementary materials).
A previously published analysis of the observed spectral energy distribution of the emission from the source reveals an appreciable stellar emission component, originating from $5.6_{-0.4}^{+2.8}\times10^{10}\,\Msun$ in stars \cite{Bongiorno2012}.
Our own analysis provides a yet lower stellar mass, of $\mstar=4.4_{-0.5}^{+0.4}\times10^{10}\,\Msun$.
However, we focus on the previously determined, higher stellar mass, as a conservative estimate.
The source is also detected at far-IR and (sub)millimeter wavelengths, which allows us to constrain the SFR in the host galaxy to about $400\,\Msol\,{\rm year}^{-1}$.
The stellar mass of the host galaxy is consistent with the typical value for star-forming galaxies at $z\sim3$ to $4$ [i.e., the ``break'' in the mass function of galaxies; \cite{Ilbert2013_UltraVISTA}]. 
Similarly, the combination of $\mstar$ and SFR is consistent with the typical values observed at $z\sim3$ to $4$, which appear to follow the so-called main sequence of star-forming galaxies \cite{Speagle2014}.
Thus, the host galaxy of \mysobj\ is a typical star-forming galaxy for its redshift, representing a population with a number density of about $5\times10^{-5}\,\NDunit$ [e.g., \cite{Ilbert2013_UltraVISTA}].
This suggests that neither the intense, ionizing radiation that emerged during the fast SMBH growth, 
nor the AGN-driven outflow, have quenched star formation in the host galaxy.
The relatively high stellar mass and SFR of the host galaxy further suggest that it is unlikely that the AGN affected the host in yet earlier epochs.
That is, even in this case of extreme SMBH growth, there is no sign of AGN-driven suppression of star formation in the host.
%

\paragraph*{}
\label{sec:discussion}

\begin{figure*}[t]
\centering
\includegraphics[width=0.95\textwidth, angle=-90]{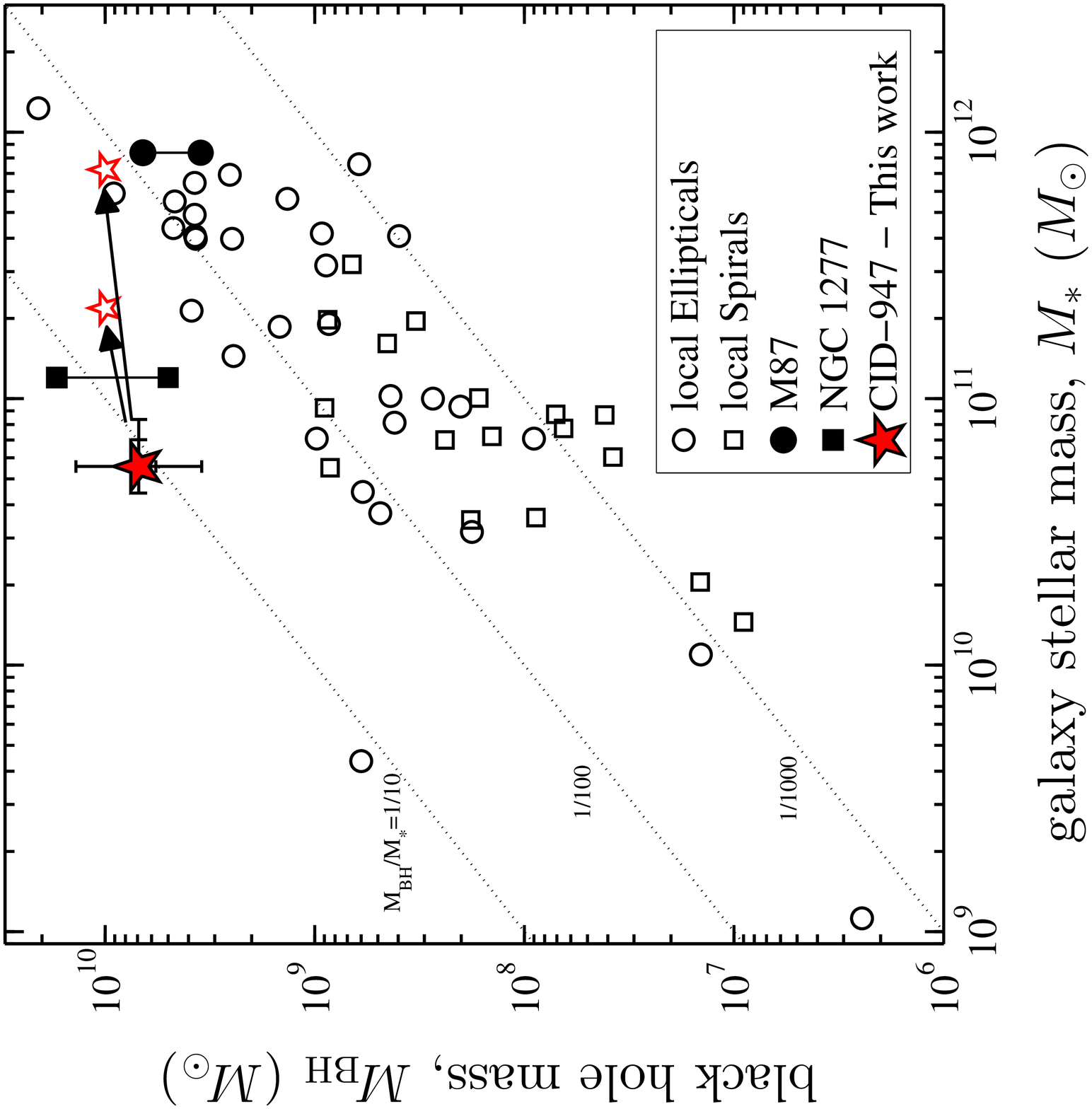} 
\caption{
{\bf A comparison of \mysobj\ with a compilation of observed $\mbh$ and $\mstar$ estimates in the local universe} 
[adapted from \cite{KormendyHo2013_MM_Rev}, assuming the tabulated bulge-to-total fractions].
\mysobj\ (red star) has a very high BH-to-stellar mass ratio of $\mbh/\mstar\simeq1/10$.
The asymmetric error bars shown on $\mbh$ and $\mstar$ represent measurement-related uncertainties, while the symmetric ones demonstrate systematic uncertainties of 0.3 dex (on $\mbh$) and 0.1 dex (on $\mgal$).
The masses inferred for subsequent growth scenarios are highlighted as empty red stars.
The \mysobj\ system is expected to evolve only mildly in $\mbh$ (perhaps to $\sim10^{10}\,\Msol$), but $\mstar$ should grow to at least $2\times10^{11}\,\Msol$, and possibly to as much as $\sim7\times10^{11}\,\Msol$, by $z=0$.
The local galaxies NGC~1277 and M87, which could be considered as descendants of systems like \mysobj, are highlighted as filled symbols [\cite{Emsellem2013_NGC1277} and \cite{Walsh2013_M87}, respectively].
Some studies suggest these galaxies to have somewhat higher $\mbh$, and therefore relatively high mass ratios, of $\mmsmall= 1/7$ and $1/127$, respectively \cite{Gebhardt2011_M87,VandenBosch2012_NGC1277}.
}
\label{fig:MM_local_evo}
\end{figure*}

\begin{figure*}[t]
\centering
\includegraphics[width=0.90\textwidth, angle=-90]{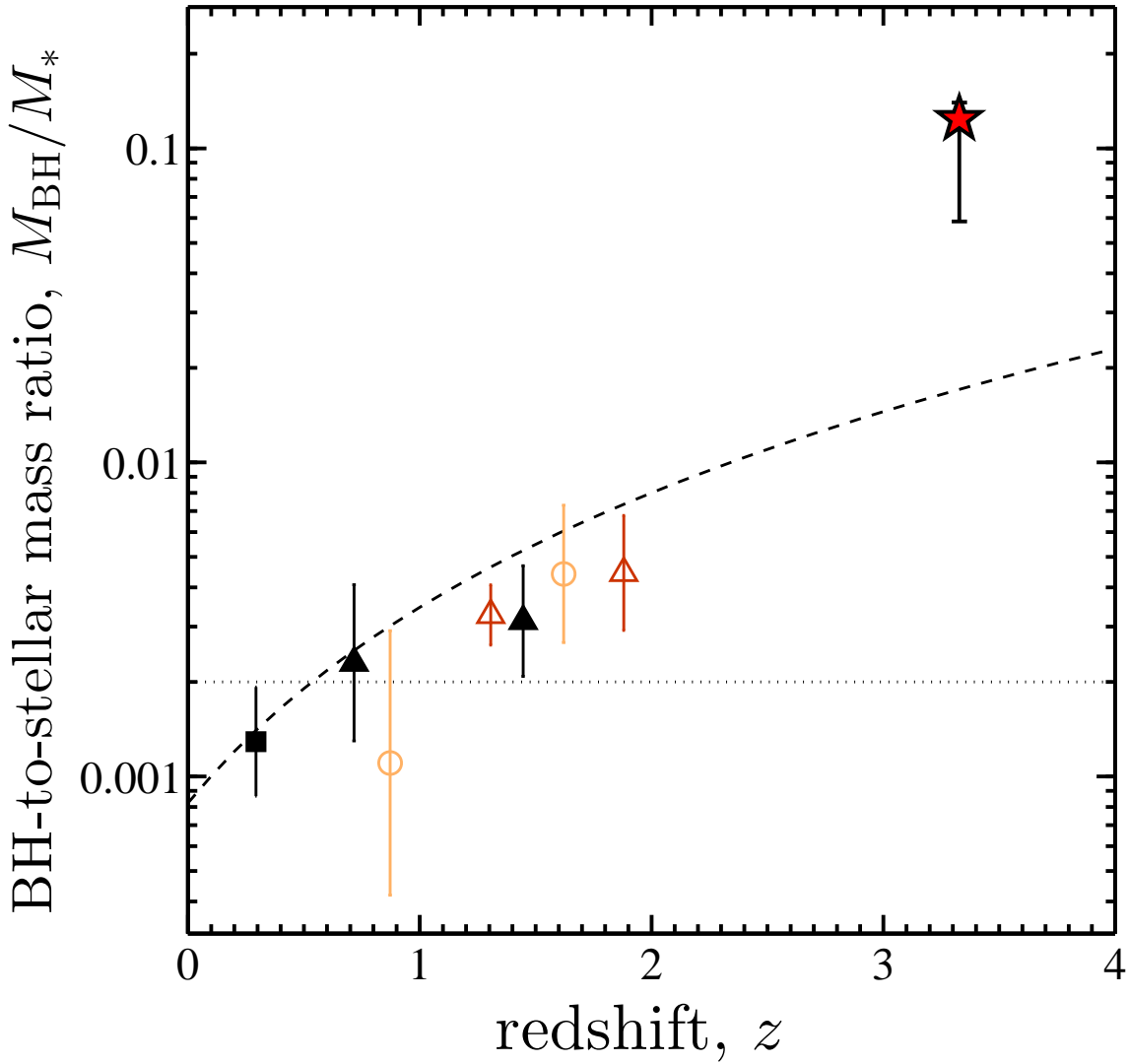} 
\caption{
{\bf The observed cosmic evolution of the BH-to-stellar mass ratio, $\mmsmall$, and its extrapolation beyond $z\sim2$.}
\mysobj\ (red star) has $\mmsmall=1/8$ at $z\simeq3.3$, which is higher by a factor of at least $\sim50$ than the typical value in local, inactive galaxies (at most, $\mmsmall \sim 1/500$; dotted line).
The error bars shown for \mysobj\ represent only the measurement-related uncertainties, propagating the uncertainties on $\mbh$ and on $\mgal$.
The different data points at $z<2$ represent typical (median) values for several samples with $\mmsmall$ estimates, with uncertainties representing the scatter within each sample [filled symbols, open circles, and open triangles represent samples from \cite{Decarli2010_MM_evo}, \cite{Peng2006_MM_lensed}, and \cite{Merloni2010}, respectively; adapted from \cite{Decarli2010_MM_evo}].
Even compared to the extrapolation of the evolutionary trend supported by these lower-redshift data, $\mmsmall\sim\left(z+1\right)^2$ [dashed line, scaled as in \cite{McLure2006_MM}], \mysobj\ has a significantly higher $\mmsmall$.
}
\label{fig:MM_vs_z}
\end{figure*}

Our analysis indicates that the BH-to-stellar mass ratio for \mysobj\ is $\mmsmall\simeq1/8$.
In comparison, most local (dormant) high-mass BHs typically have $\mmsmall\sim1/700$ to $1/500$ [see Fig.~\ref{fig:MM_local_evo} and, e.g., \cite{HaringRix2004,KormendyHo2013_MM_Rev}]. 
The $\mmsmall$ value that we find for \mysobj\ is thus far higher than typically observed in high-mass systems in the local universe, by at least an order of magnitude and more probably by a factor of about 50.
The only local system with a comparably extreme mass ratio is the galaxy NGC~1277, which was reported to have $\mmsmall\simeq1/7$ [with $\mbh=1.7\times10^{10}\,\Msol\simeq2.5\times\mbh$(\mysobj); see \cite{VandenBosch2012_NGC1277}, but also \cite{Emsellem2013_NGC1277}].
At earlier epochs (still $z<2$), the general trend is for $\mmsmall$ to increase slightly with redshift, but typically not beyond $\mmsmall\sim1/100$ (see Fig.~\ref{fig:MM_vs_z}).
Only a few systems with reliable estimates of $\mbh$ show $\mmsmall$ reaching as high as $1/30$ [e.g., \cite{Decarli2010_MM_evo,Merloni2010,Bennert2011}].

Given the high masses of both the SMBH and stellar population in \mysobj, we expect this system to retain an extreme $\mmsmall$ throughout its evolution, from $z=3.328$ to the present-day universe.
Because the $\mbh$ that we find is already comparable to the most massive BHs known, it is unlikely that the SMBH will experience any further appreciable growth (i.e., beyond $\mbh\simeq10^{10}\,\Msol$).
Indeed, if the SMBH accretes at the observed rate through $z=2$, it will reach the extreme value of $\sim 10^{10}\,\Msol$, and by $z=1$ it will have a final mass of $\sim2.5\times10^{10}\,\Msol$.
As for the host galaxy, we can constrain its subsequent growth following several different assumptions.
First, if one simply assumes that the galaxy will become as massive as the most massive galaxies in the local universe [$\mgal\simeq10^{12}\,\Msol$; \cite{Baldry2012_GAMA_MF}], then the implied final mass ratio is on the order of $\mmsmall\sim1/100$.
Alternatively, we consider more realistic scenarios for the future growth of the stellar population, relying on the observed mass ($\mgal$) and growth rate (SFR).
Our calculations involve different scenarios for the decay of star formation in the galaxy (see section \ref{SM_sec_mstar_evo} in the supplementary materials), 
and predict final stellar masses in the range $\mgal\left(z=0\right)\simeq \left(2-7\right)\times10^{11}\,\Msol$, 
which is about an order of magnitude higher than the observed mass at $z=3.328$.
The inferred final mass ratio is $\mmsmall\sim1/50$.
This growth can only occur if star formation continues for a relatively long period ($\gtsim1$ Gy) and at a high rate ($>50\,\Msol\,{\rm year}^{-1}$).
This would require the presence of a substantial reservoir, or the accretion, of cold gas, which, however, could not increase the SMBH mass by much.
Finally, in the most extreme scenario, the star formation shuts down almost immediately (i.e., due to the AGN-driven outflow), and 
the system remains ``frozen'' at $\mmsmall\sim1/10$ throughout cosmic time.
If the SMBH does indeed grow further (i.e., beyond $10^{10}\,\Msol$), this would imply yet higher $\mmsmall$.
Thus, the inferred final BH-to-stellar mass ratio for \mysobj\ is, in the most extreme scenarios, about $\mmsmall\sim1/100$, and probably much higher (see Fig.~\ref{fig:MM_local_evo}).

\mysobj\ therefore represents a progenitor of the most extreme, high-mass systems in the local universe, like NGC~1277.
Such systems are not detected in large numbers, perhaps due to observational selection biases.
The above considerations indicate that the local relics of systems like \mysobj\ are galaxies with at least $\mgal\sim5\times10^{11}\,\Msol$. 
Such systems are predominantly quiescent (i.e., with low star-formation rates, ${\rm SFR} \ll 1\,\Msol\,{\rm year}^{-1}$) and relatively rare in the local universe, with typical number densities on the order of $\sim10^{-5}\,\NDunit$ \cite{Baldry2012_GAMA_MF}.
We conclude that \mysobj\ provides direct evidence that at least some of the most massive BHs, with $\mbh\gtsim10^{10}\,\Msol$, already in place just 2 Gy after the Big Bang, did not shut down star formation in their host galaxies. 
The host galaxies may experience appreciable mass growth in later epochs, 
without much further black hole growth, 
resulting in very high stellar masses but still relatively high $\mmsmall$.
Lower-mass systems may follow markedly different coevolutionary paths.
However, systems with $\mmsmall$ as high as in \mysobj\ may be not as rare as previously thought, as they can be consistently observed among populations with number densities on the order of $\sim10^{-5}\,\NDunit$, both at $z>3$ and in the local universe, and not just among the rarest, most luminous quasars.




\bibliographystyle{Science}

\section*{Acknowledgments}

The new MOSFIRE data presented herein were obtained at the W.M. Keck Observatory, which is operated as a scientific partnership among the California Institute of Technology, the University of California and the National Aeronautics and Space Administration. 
The Observatory was made possible by the generous financial support of the W.M. Keck Foundation.
We thank M. Kassis and the rest of the staff at the W.M. Keck observatories at Waimea, HI, for their support during the observing run.
We recognize and acknowledge the very significant cultural role and reverence that the summit of Mauna Kea has always had within the indigenous Hawaiian community.  
We are most fortunate to have the opportunity to conduct observations from this mountain.
Some of the analysis presented here is based on data products from observations made with European Southern Observatory (ESO) Telescopes at the La Silla Paranal Observatory under ESO program ID 179.A-2005 and on data products produced by TERAPIX and the Cambridge Astronomy Survey Unit on behalf of the UltraVISTA consortium.
We are grateful to A. Faisst and M. Onodera for their assistance with the acquisition and reduction of the MOSFIRE data. 
We thank S. Tacchella, J. Woo, and W. Hartley for their assistance with some of the evolutionary calculations.
K.S. gratefully acknowledges support from Swiss National Science Foundation Professorship grant PP00P2\_138979/1. 
F.C. acknowledges financial support by the NASA grant GO3-14150C.
M.E. acknowledges financial support by the NASA \chandra\ grant GO2-13127X.
B.T. is a Zwicky Fellow at the ETH Zurich.

\vspace*{2cm}

\noindent
The Supplementary Materials, available on Science Online, include:\\
Data, Methods, and Supplementary Text S1 to S5\\ 
Figs. S1 to S4\\
Table S1\\
References ({\it 31-81})\\

\clearpage

\clearpage

\clearpage

\paragraph*{}

\makeatletter 
\renewcommand{\thefigure}{S\@arabic\c@figure}
\makeatother

\makeatletter 
\renewcommand{\thetable}{S\@arabic\c@table}
\makeatother

\makeatletter 
\renewcommand{\thesection}{S\@arabic\c@section}
\makeatother

\setcounter{section}{0}
\setcounter{figure}{0}
\setcounter{table}{0}

\begin{center}
  {\Large Supplementary Materials for}

  \vspace{0.5cm}
  \textbf{\large An Over-Massive Black Hole in a Typical Star-Forming Galaxy, 2 Billion Years After the Big Bang}

  \hfill
  \begin{minipage}{\textwidth}
    \begin{center}
      \vspace{0.5cm}
      \textbf{Authors:}
      Benny Trakhtenbrot,$^{\ast}$\\
      C. Megan Urry,
      Francesca Civano,
      David J. Rosario,
      Martin Elvis,
      Kevin Schawinski,
      Hyewon Suh,
      Angela Bongiorno,
      Brooke D. Simmons\\
    \end{center}
  \end{minipage}

  \vspace{0.5cm}
  \normalsize{$^\ast$To whom correspondence should be addressed; E-mail: benny.trakhtenbrot@phys.ethz.ch}
\end{center}

\vspace{2cm}
\textbf{This PDF file includes:}

\hfill
\begin{minipage}{\dimexpr\textwidth-2.0cm}
  \vspace{0.5cm}
  Data, Methods and Supplementary Text \S\S S1-S5\\
  Figures S1 to S4 \\
  Table S1 \\
  Additional references 31 to 81\\
\end{minipage}

\clearpage

In these Supplementary Materials, we provide additional details about all the aspects of our study presented and discussed in the main article, including:  
the new Keck/MOSFIRE $K$-band observations and their analysis (\S\ref{SM_sec_Kband_spec});
the multiwavelength data and related derivation of stellar mass ($\mgal$) and star formation rate (SFR; \S\ref{SM_sec_SED}); 
the derivation of black hole mass ($\mbh$) and accretion rate ($\lledd$), and a discussion of the SMBH evolution (\S\ref{SM_sec_M_LLedd}); 
the rest-frame UV spectrum, broad absorption features and properties of the AGN-driven outflow (\S\ref{SM_sec_BAL}); 
and of the calculations of possible final $\mgal$ of the host galaxy (\S\ref{SM_sec_mstar_evo}).

\section{New Keck/MOSFIRE $K$-band Data and Analysis}
\label{SM_sec_Kband_spec}

\subsection{Keck/MOSFIRE Observations and Data Reduction}
\label{SM_sec_obs_red}

The source \mysobj\ (J2000.0 coordinates $\alpha=$ 10:01:11.35, $\delta=$ +02:08:55.6) was observed with the Keck/MOSFIRE instrument \cite{McLean2012_MOSFIRE} during the night of January 23-24th., 2014, with observing time allocated through the Yale-Caltech collaborative agreement. 
We used the normal \kband\ setup, which covers order 4 of the 110.5 mm$^{-1}$ reflection grating. 
As \mysobj\ was our primary target, it was positioned near the center of the mask, providing a spectral coverage of $\lambda=$19,415--23,837 \AA.
To ensure adequate coverage of the sky background emission, and its subtraction from the AGN signal, we used 4 pairs of MOSFIRE bars, to form a 24\arcsec-long pseudo-slit. The MOSFIRE pixel scale is 0.18\arcsec/pix.
To prevent significant slit losses, we set the slit width(s) to 1\arcsec, which resulted in a spectral resolution of about $R\equiv\lambda/\Delta\lambda=3600$.
Observational conditions during the night were generally good, with typical seeing of $\sim$0.8\arcsec\ in the $K$-band during the science observations.
The science exposures, totaling an hour, consisted of 20 separate sub-exposures of 3 minutes each, dithered between two positions along the slit with a separation of 4\arcsec, to allow for an accurate subtraction of the sky emission.
The typical airmass during the observations was about 1.08.
Several times during the night we also observed the A0v stars HIP-34111 and HIP-56736, as well as the fainter white dwarf GD71, to allow a robust flux calibration. 

The data were reduced using a combination of different tools. 
First, we used the dedicated MOSFIRE pipeline (2014.06.10 version) to obtain flat-fielded, wavelength calibrated 2D spectra of all the sources observed within each mask (including the standard stars). 
The wavelength calibration was performed using sky emission lines, and the best-fit solutions achieved an rms of $\sim0.1$\AA.
Next, we used standard {\tt IRAF} procedures to produce a 1D spectrum, using an aperture of 11 pix (i.e., 2\arcsec).
Finally, we used the {\tt Spextool} IDL package to remove the telluric absorption features near 2 \mic\ and to perform the relative and absolute flux calibrations, based on a detailed library spectrum of Vega \cite{Vacca2003,Cushing2004}.
The absolute flux calibration we obtained is in excellent agreement with the archival photometry available for \mysobj:
the synthetic magnitude derived from the spectrum is $K_{\rm s}=20.03$ (AB magnitudes), compared with the archival value of $20.00\pm0.01$ \cite{McCracken2012_COSMOS_UltraVISTA}. 
We however chose to apply the minor scaling needed to match the archival photometry (a factor of 1.03), in order to be fully consistent with the value used in the SED decomposition (\S\ref{SM_sec_SED}).
We finally note that \mysobj\ is one of several COSMOS targets observed in this campaign, and we verified the robustness of the different reduction steps by visually verifying that the same reduction yields artifact-free spectra for the other sources. 
The typical signal-to-noise across the core part of the spectrum is $S/N\sim5-7$ per pixel. 
After re-binning the spectrum to a uniform spacing of 1 \AA\ (in rest-frame; $60\,\kms$), we obtain $S/N\sim7-10$ per spectral resolution element.

\subsection{Spectral Analysis of \Hbeta\ Emission Complex}
\label{SM_sec_analysis}

We modeled the \kband\ spectrum of \mysobj\ to measure the monochromatic continuum luminosity at (rest-frame) wavelength of 5100\AA\ ($\lamLlam\left[5100{\rm \AA}\right]$, or $\Lop$) and the width of the broad \Hbeta\ emission line ($\fwhb$).
The analysis methodology is very similar to that discussed in numerous previous works (e.g., \cite{Netzer_Trakht2007,Shen_dr7_cat_2011,TrakhtNetzer2012_Mg2}, and references therein), and is only briefly described here.

\begin{figure*}[t!]
\centering
\includegraphics[width=0.75\textwidth]{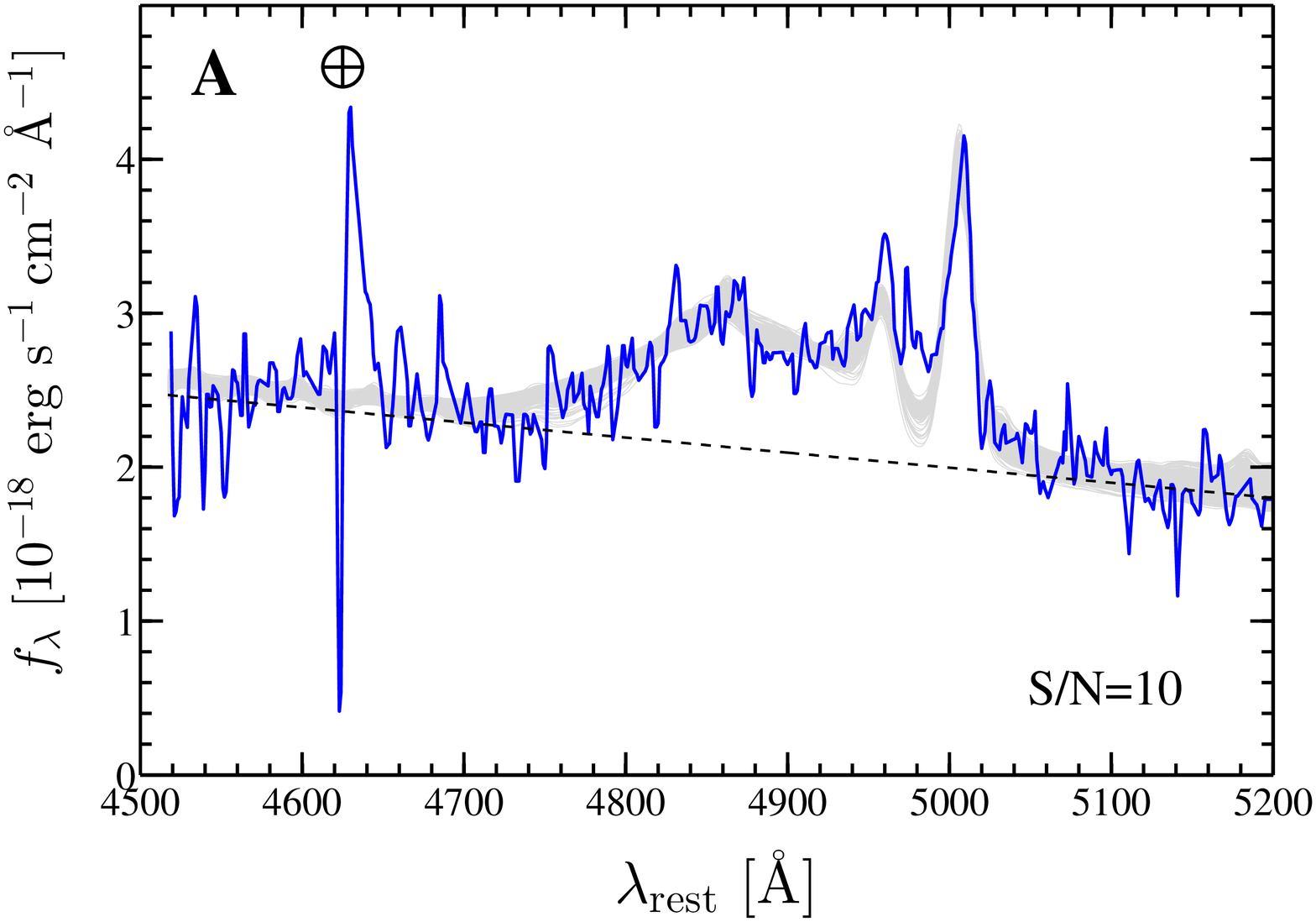} 
\includegraphics[width=0.42\textwidth, angle=-90]{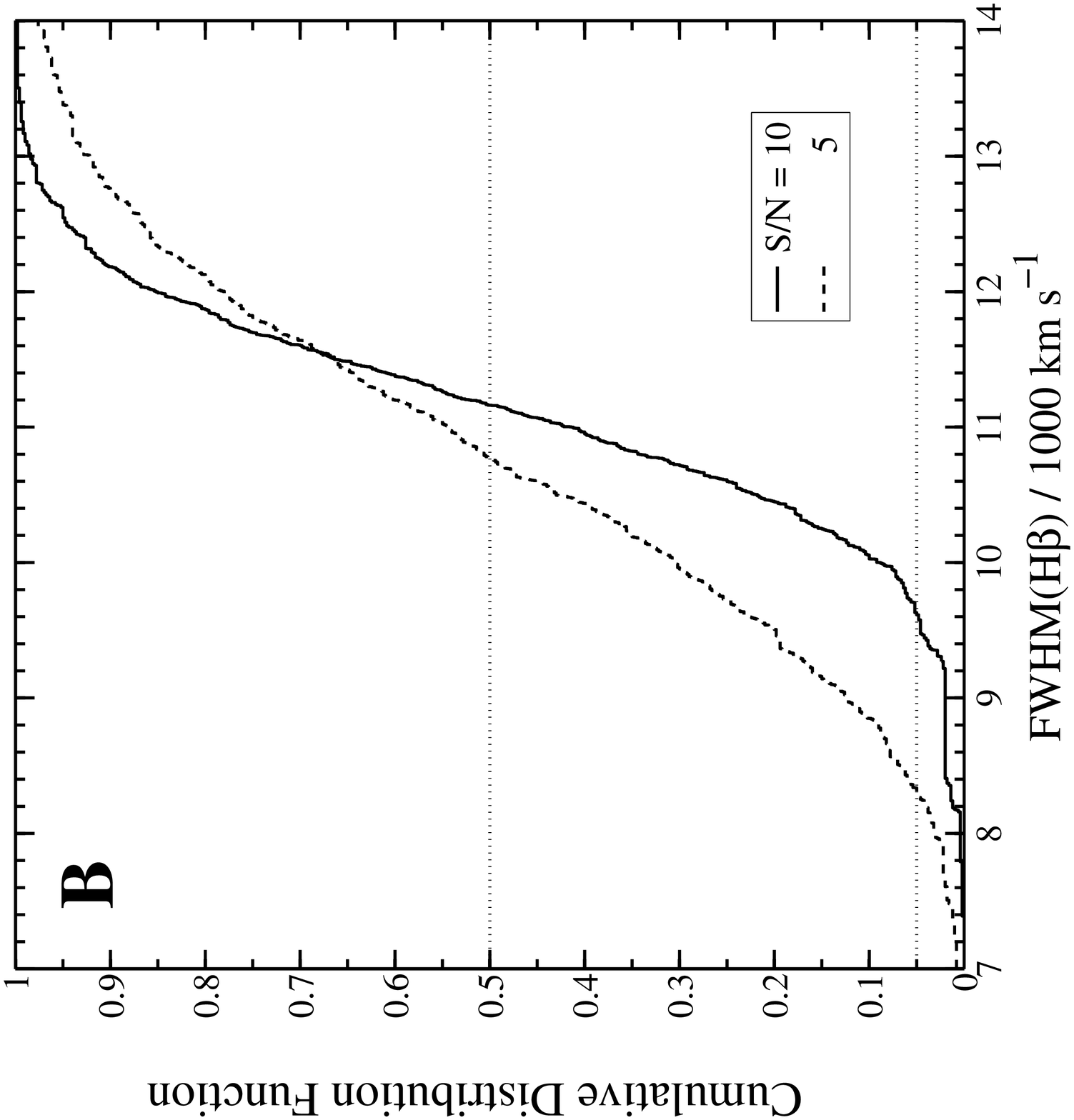} 
\includegraphics[width=0.42\textwidth, angle=-90]{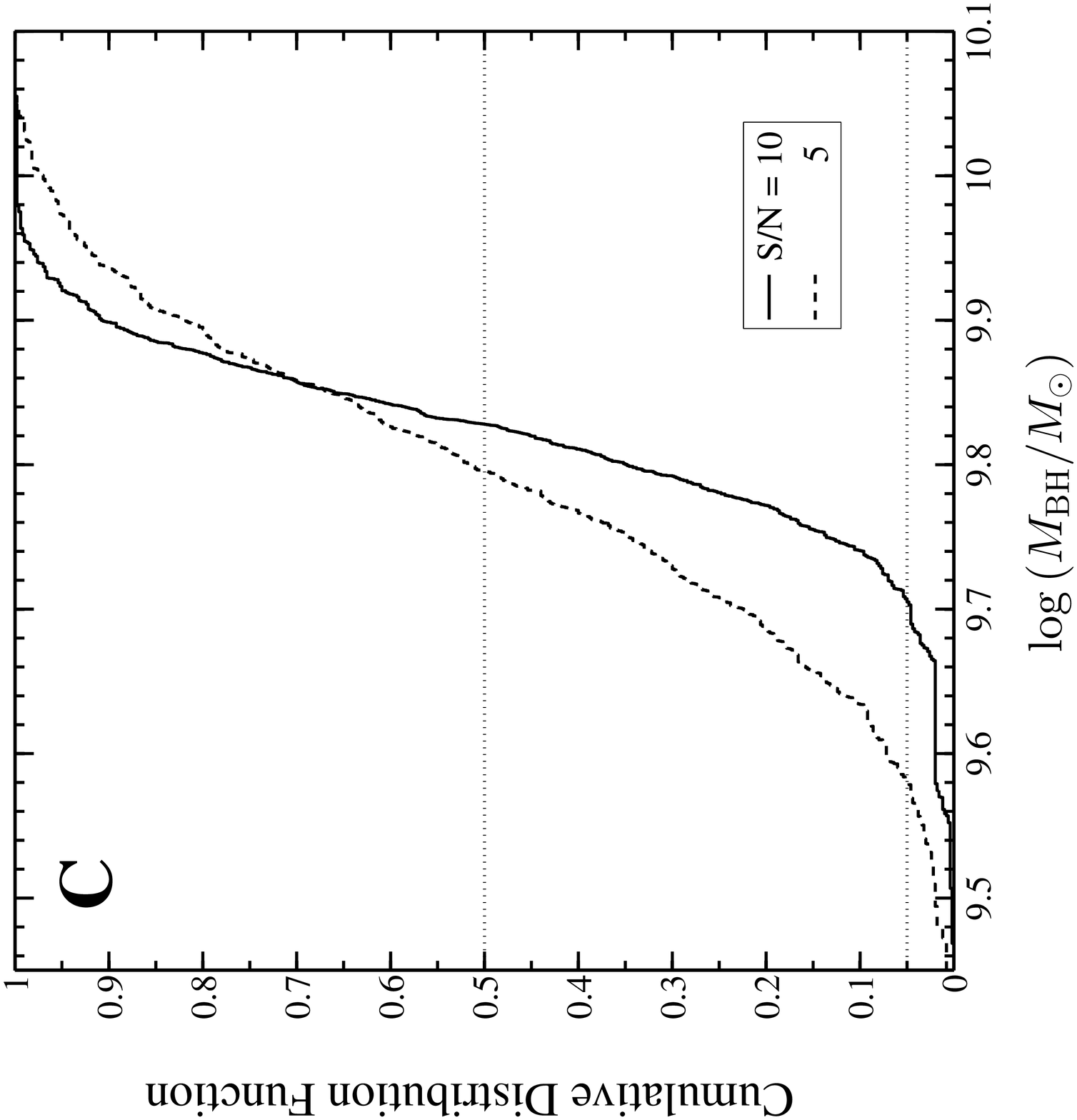} 
\caption{
Results of the re-sampling simulations for the \Hbeta\ emission complex of \mysobj. 
\emph{Top} - ({\bf A}): Results for 500 simulations, with $S/N=10$ and ``nominal'' \oiii\ profiles (i.e., single-Gaussian).
The diagram illustrates the original spectrum (blue), the best-fit linear continuum (dashed black), and the collection of 500 models that fit the simulated spectra (gray shaded region).
\emph{Bottom:} Cumulative distribution functions (CDFs) of the obtained line widths, $\fwhb$ ({\bf B}) and SMBH masses, $\mbh$ ({\bf C}).
In both panels, the solid lines illustrate the results for the $S/N=10$ simulation, while the dashed lines correspond to a more conservative simulation, with   $S/N=5$.
The (dotted) horizontal lines mark the 5 and 50\% (i.e., median) levels. 
We find that 95\% of the $S/N=10$ simulations resulted in $\fwhb>9615\,\kms$ and $\mbh>5.09\times10^{9}\,\Msol$; and that 99\% resulted in  $\fwhb>8175\,\kms$ and $\mbh>3.61\times10^{9}\,\Msol$. 
}
\label{fig:CID_947_sim}
\end{figure*}

The spectra were modeled with a linear (pseudo) continuum, a broadened \feii\ template \cite{BG92}, and a combination of Gaussians which account for the broad and narrow emission lines, namely \Hbeta, [O\,{\sc iii}]\,$\lambda\lambda$4959,5007, and \HeIIop.
The \Hbeta\ model consists of a broad component (modeled with 2 Gaussians) and a narrow component, which is tied to the width of the \oiii\ lines.
The continuum flux at 5100\AA\ was estimated directly from the best-fit linear continuum.\footnote{The monochromatic luminosity, as all other luminosities and ages reported here, are calculated assuming a cosmological model with $\Omega_{\Lambda}=0.7$, $\Omega_{\rm M}=0.3$, and $H_{0}=70\,\kms\,{\rm Mpc}^{-1}$.}
We preferred to use $\fwhm$ over $\sigBLR$ as the probe of the virial velocity field of the BLR gas, as the former can be more robustly estimated in spectra of limited S/N, as is the case with our MOSFIRE data \cite{Denney2013_low_SNR}. 
However, since the best-fitting model for \Hbeta\ turned out to be overwhelmingly dominated by a single broad Gaussian component (the flux ratio between the two components is 45:1), 
the differences between the results derived by two approaches are expected to be negligible.
The best-fit models are presented in Figure~\ref{fig:CID_947_sim}.
The relevant best-fit parameters resulting from our fitting of the MOSFIRE spectrum are 
$\fwhb=11330 \,\kms$, and
$\Lop=4.16\times10^{45}\,\ergs$.
After accounting for host-galaxy contamination of about 14\% (following the analysis presented in \S\ref{SM_sec_SED}), the intrinsic optical luminosity becomes $\Lop=3.58\times10^{45}\,\ergs$.
We note that the broad \Hbeta\ profile in our best-fit model is highly symmetric, with an asymmetry index of ${\rm A.I.}=0.03$, consistent with the typical value found in large samples of un-obscured AGN (see, e.g., \cite{Sulentic2000_BLR_rev}).
The broad component is, however, blue-shifted by about $830\,\kms$, relative to the expected wavelength (at the source systemic redshift).
Such large blue-shifts are relatively rare, with an occurrence rate of only about 5\%.\footnote{This estimate is based on the \Hbeta\ measurements of about 20,000 SDSS AGN, presented in Ref. \cite{TrakhtNetzer2012_Mg2}.}
To verify that our estimate of $\fwhb$ is not severely affected by these properties of broad \Hbeta\ profile, we obtained an alternative estimate of the line width, which relies only on the blue part of the profile, which is not blended with Iron and \oiii\ emission.
This alternative estimate is obtained by doubling the one-sided line width, which was measured following the approach described in Ref. \cite{Peterson2004}.
This results in a $\fwhb$-equivalent of $9236\,\kms$, only 0.09 dex lower than our best-fit value. 
However, we stress that this alternative estimate is far less robust than our fiducial, best-fit value, as it relies on the identification of the (spectral) pixel with maximal flux density, which is very sensitive to small-scale flux density fluctuations in spectra of limited S/N, as in the present case.
We conclude that our estimate of the width of the broad component of \Hbeta\ is not significantly affected by the shape or shift.

Due to the complexity of the spectral fitting procedure, the best approach to derive the measurement-related uncertainties on $\Lop$ and $\fwhb$ (and therefore on $\mbh$; see \S\ref{SM_sec_M_LLedd} below) is via re-sampling of the data. 
To this end, we constructed a set of 500 artificial spectra, by adding normally-distributed random noise to the observed spectra of \mysobj, scaled to provide either $S/N=10$ (comparable to the noise level in the observed data). 
An additional set of 500 simulations assumed a more conservative noise level of $S/N=5$, to verify that our results are not driven by an under-estimation of the noise level in the data.
We then re-fitted each of these artificial spectra, using the same fitting procedure as described above. 
The resulting best-fitting models are illustrated in the top panel of Figure~\ref{fig:CID_947_sim}.
After measuring $\Lop$ and $\fwhb$ for each of these best-fitting models, we obtained an artificial sample of $\fwhb$ values, the cumulative distribution function of which is illustrated in the lower-left panel of Fig.~\ref{fig:CID_947_sim}.
Fig.~\ref{fig:CID_947_sim} clearly demonstrates that, even under the conservative assumption of $S/N=5$, about 95\% of our simulations resulted in $\fwhb\gtsim8600\,\kms$.
This conservative lower limit on $\fwhb$ is lower than the best-fit value by about 0.14 dex.

The spectral region adjacent to the \oiii\ lines may suggest that their profiles may include an additional broad component (i.e., a ``blue wing'').
We therefore performed yet another set of simulations, with an alternative version of the fitting procedure that allows for an additional broad component for \oiii. 
The broad components for the two \oiii\ lines were forced to share a common width (in the range $500-1400\,\kms$) and relative shift (in the range $-350-\,+150,\kms$).
These limits are motivated by the distributions of line widths and shifts found for large samples of un-obscured AGN  (e.g., \cite{Zamanov2002,Boroson2005_OIII,Komossa2008}).
The results of the simulations indicate that this adjustment to the \oiii\ profiles does not systematically affect our estimate of $\fwhb$.
The median value obtained in the simulations is $\fwhb=10150\,\kms$, and 95\% of the simulations resulted in $\fwhb>6740\,\kms$.

\section{Broad-band Spectral Energy Distribution and Estimates of $\Lbol$, $\mstar$ and SFR}
\label{SM_sec_SED}

We used the available multiwavelength data for \mysobj\ to determine the bolometric luminosity ($\Lbol$) of the AGN in \mysobj, and the stellar mass ($\mstar$) and star formation rate (SFR) of the host galaxy.
The broad-band spectral energy distribution (SED) for \mysobj\ includes data from a large variety of surveys of the COSMOS field, including data in the X-ray (\chandra\ and \xmm), optical-to-near-IR (Subaru and CFHT), mid-to-far-IR (\spitzer\ and \herschel)), and millimeter (JMCT) regimes.

The broad-band SED of \mysobj\ was analyzed in previous studies of COSMOS AGN, which derived and reported estimates of $\Lbol$.
One such analysis, based on \chandra\ X-ray data \cite{Elvis2012_SED}, yields $L_{\rm bol,\, SED}=1.31\times10^{46}\,\ergs$ (correcting for our adopted cosmology).
\footnote{Here we use the luminosity integrated between 40 \kev\ and 1 \mic, to avoid double-counting the re-processed mid-IR emission (which would add about 17\% to $\Lbol$ for \mysobj).}
Another study, based on \xmm\ X-ray data (ref.\ \cite{Lusso2011_Lbol}; XMM-ID 60131), gives $L_{\rm bol,\, SED}=1.81\times10^{46}\,\ergs$ (also cosmology-corrected).
The difference between these two values, of a factor of 1.6, is mostly due to the markedly different X-ray fluxes reported for \mysobj\ in the \chandra\ and \xmm\ surveys of the COSMOS field, 
which may be due to intrinsic source variability.
We note that the stellar component in the SED (see below) has a negligible contribution to these estimates of $\Lbol$ (i.e., $<1\%$).

To derive the host galaxy properties, the SED of \mysobj\ was modeled separately for the rest-frame UV-optical-NIR, and for the Mid-IR-to-millimeter regime.

\begin{figure}[t]
\begin{center}
\centering
\includegraphics[width=0.46\textwidth]{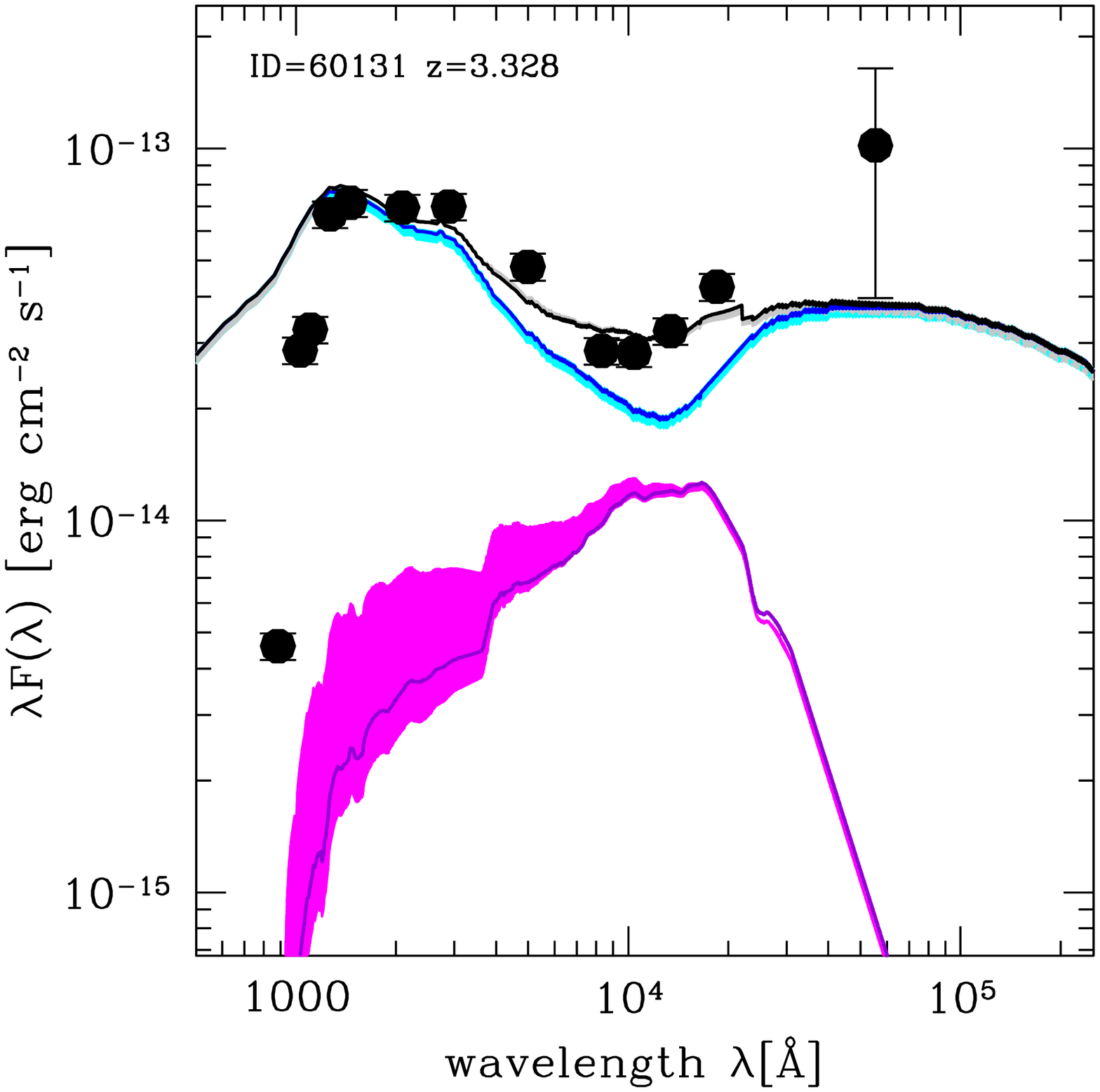}
\includegraphics[width=0.53\textwidth]{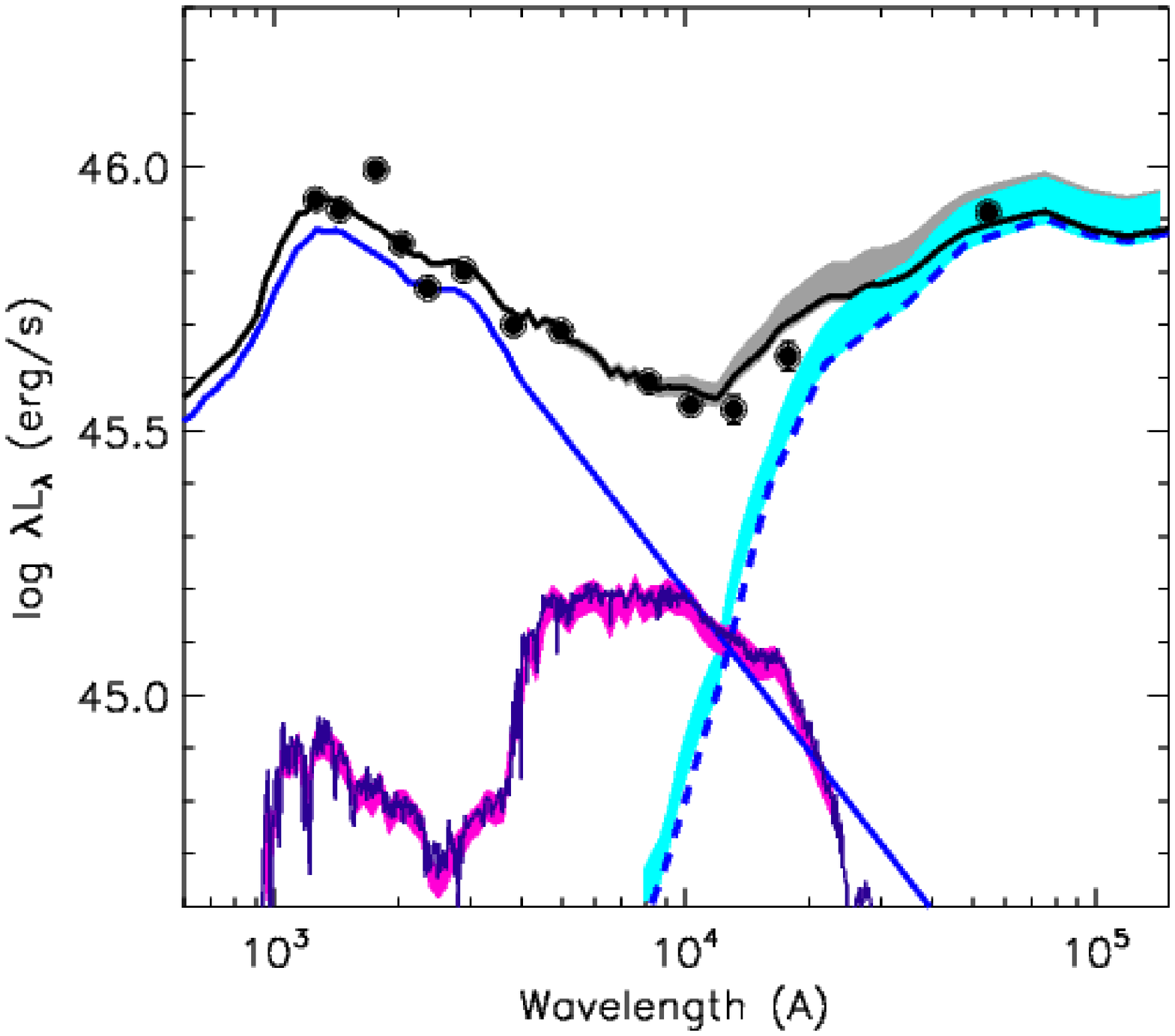}
\caption{
The UV-to-IR SED of \mysobj, based on the available ground-based and \spitzer\ imaging data in the COSMOS field, which is used to infer the stellar mass of the host galaxy.
\emph{Left} - SED fitting from Bongiorno et al.\ (2012; \cite{Bongiorno2012}). 
The observed SED (black points) is fit by a model (black line) that consists of three distinct components:
an un-obscured AGN, dominating the UV regime (solid blue)
and a stellar population (magenta), which contributes a significant fraction of the emission around rest-frame wavelength of $\sim1$ \mic.
\emph{Right} - our own best-fitting model, in which the dusty toroidal structure, dominating the mid-IR regime (dashed blue), is treated separately from the intrinsic AGN emission. 
}
\label{fig:cid947_opt_SED}
\end{center}
\end{figure}

The rest-frame UV-optical-NIR part of the SED includes the emission from the accreting SMBH, a part of which is re-processed by a dusty toroidal structure (``torus'') and re-emitted in IR wavelengths, and from stellar population of the host galaxy.
The data in this regime consists of flux measurements in 13 spectral bands (obtained with Subaru, CFHT and \spitzer), ranging from $\sim3700$ \AA\ (CFHT/$u^*$) to 24 \mic\ (\spitzer/MIPS).
We first rely on the data accumulated and analyzed in a previous COSMOS study by Bongiorno et al.\ \cite{Bongiorno2012}.
Here we mention briefly only some of the features of this modeling, and refer the reader to \cite{Bongiorno2012} for a detailed discussion.
In that study, the data were modeled as the sum of two distinct components, representing the emission originating from the AGN and from stars in the host galaxy. 
The AGN component is described by the multiwavelength AGN SED of Richards et al.\ (2006; \cite{Richards2006_SED}).
The stellar component was described by a grid of models, produced by a well-established stellar popular synthesis procedure \cite{Bruzual2003}.
Each of the templates represents a stellar population with a different age (ranging from 50 Myr - 1.88 Gyr) and exponential decay rate ($\tau_{\rm SFH}=0.1-30$ Gyr), and further assume a Chabrier initial mass function \cite{Chabrier2003_IMF_rev}. 
The templates were then subjected to both nuclear and galaxy-wide dust extinction (with $E_{\rm B-V}$ values of up to 1 and 0.5, respectively).
The best-fit model for \mysobj\ in the Bongiorno et al.\ study (Fig.~\ref{fig:cid947_opt_SED}, left) provides a stellar mass of $\mgal=5.57^{+2.78}_{-0.38} \times10^{10}\,\Msol$, with a reddening of $E_{\rm B-V}=0.5$.
The fraction of the total monochromatic luminosity at 5100 \AA\ which is contributed by the stellar component is about $f_{\rm host}\left(5100\right)=0.14$. 
This host contamination is taken into account when we estimate the mass of the SMBH (see \S\ref{SM_sec_M_LLedd}).

We have repeated the UV-to-IR SED fitting, using the most up-to-date imaging data available for \mysobj\ (UltraVISTA DR2, \cite{McCracken2012_COSMOS_UltraVISTA}; see Table~\ref{tab:sed}), and a slightly modified AGN model, in which the emission from the dusty torus (dominating the mid-IR regime) is separated from the intrinsic AGN radiation (dominating the UV-optical regime).
Our analysis resulted in a very similar stellar mass to the aforementioned one.
The stellar component is represented, as before, by the Richards et al.\ AGN SED, which is however extrapolated as a power-law at $\lambda_{\rm rest}>4000$ \AA. 
The IR emission from the dusty torus is represented by composites from a dedicated study of the IR SEDs of AGN \cite{Silva2004}.
The grid of stellar population models have remained the same as in the aforementioned ``reference'' SED fit, with the age of the stellar population capped at the age of the Universe at $z=3.328$.
Our choice of the Bruzual \& Charlot models is motivated by the fact that they were also used in most studies of star forming galaxies at $z>3$, which we use here as reference (e.g., \cite{Ilbert2013_UltraVISTA,Speagle2014}).
For a detailed discussion of the effects of alternative stellar population models, e.g., \cite{Ilbert2013_UltraVISTA} and \cite{Speagle2014}. 
We have restricted the components so that the UV-optical part of the SED would be dominated by the AGN, i.e. $f_{\rm AGN}>f_{\rm host}$.
This is motivated by the overall AGN luminosity of \mysobj\ (see above) and the fact that the rest-frame UV spectrum does not show significant host contamination in the deep absorption features (see \S\ref{SM_sec_BAL}). 
We have also explicitly omitted the data below $\lambda_{\rm rest}=1216$ \AA\ (i.e., the $u$ and $B$ bands), as these are expected to be affected by \Lya\ absorption by the intergalactic medium along the line of sight to \mysobj. 
The resulting additional best-fit model (Fig.~\ref{fig:cid947_opt_SED}, right) relies on a stellar population with an age of 1 Gyr, a stellar mass of $\mgal=4.37^{+0.42}_{-0.48} \times10^{10}\,\Msol$, with a reddening of $E_{\rm B-V}=0.05$.
In this new fit, the stellar component contributes $f_{\rm host}\left(5100\right)=0.36$.
Next, we re-fitted the data with a restricted model in which the stellar population is kept at the oldest reasonable age (1.8 Gyr). 
Since older stellar populations have higher mass-to-light ratios, such a fit would in principle provide a conservative upper limit on the stellar mass.
This fit resulted in $\mgal=6.48^{+0.28}_{-0.56} \times10^{10}\,\Msol$, and $f_{\rm host}\left(5100\right)=0.35$.
Finally, we have repeated this oldest-population fit, this time without the restriction of $f_{\rm AGN}>f_{\rm host}$. 
This resulted in $\mgal=5.95 \pm 0.21 \times10^{10}\,\Msol$, and $f_{\rm host}\left(5100\right)=0.55$.
We stress however that these latter age-restricted models do \emph{not} provide the best fits of the data, as the resulting $\chi^2$ is higher than that found for the non-restricted case.
We conclude that the best-fit stellar mass that we obtain for the host of \mysobj\ is $\mgal=4.37 \pm ^{+0.42}_{-0.48} \times10^{10}\,\Msol$.
We however choose to base the rest of the analysis on the slightly higher mass found in the Bongiorno et al.\ study, as it represents a more conservative choice given the extremely high BH-to-stellar mass ratio we find for \mysobj, and since it is based on the same SED decomposition code that was used in some of the reference studies to which we compare our results \cite{Merloni2010}.

\begin{figure}[t]
\begin{center}
\centering
\includegraphics[width=0.95\textwidth]{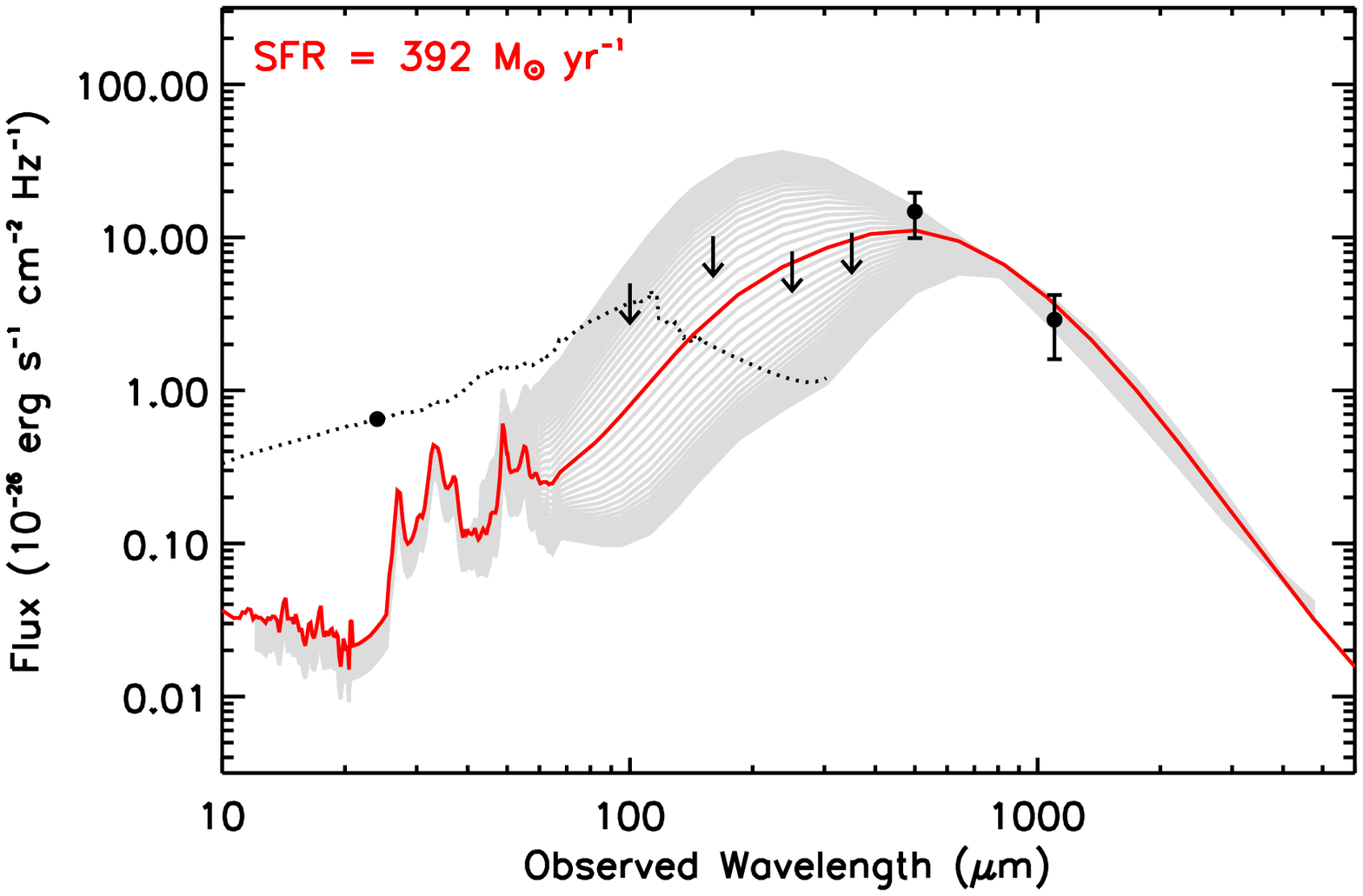}
\caption{
The mid-to-far IR SED of \mysobj, based on the available \spitzer\ data and the low resolution \herschel\ and millimeter-wave observations, which is used to infer the star formation rate of the host galaxy.
The gray lines represent a subset of the far-IR templates of star-forming galaxies we used \cite{DaleHelou_2002_SED}, with the best-fit template and the corresponding SFR highlighted in red.
The AGN contamination at sub-millimeter-to-millimeter wavelengths ($\lambda_{\rm rest} > 200\,{\rm \mu m}$) is negligible, as illustrated by the pure-AGN spectral energy distribution (black dotted line; adapted from ref.\ \cite{Netzer2007_QUEST}). 
}
\label{fig:cid947_sed}
\end{center}
\end{figure}

The Mid-IR-to-millimeter part of the SED is dominated by emission from the (cold) dusty gas in the host galaxy, heated by the star formation activity.
Here we rely on \herschel\ and JMCT/AzTEC detections at 500 \mic\ and 1.1 millimeter, and upper limits at 100, 160, 250 and 350 \mic, from \herschel\ \cite{Lutz2011_PEP,Aretxaga2011_AzTEC}.
The data were fit with a grid of dust-emission templates of star-forming galaxies, covering a representative range of SED shapes (i.e., effective temperatures; \cite{DaleHelou_2002_SED}).
The best-fit template for the far-IR and millimeter data implies a star formation rate of ${\rm SFR}=392\,\mpyr$.
We present the Mid-IR-to-millimeter data and models in Fig.~\ref{fig:cid947_sed}.
We stress that the AGN contribution to the emission in this regime is negligible, as demonstrated by the dotted black line in Fig.~\ref{fig:cid947_sed} \cite{Netzer2007_QUEST}.
The SFR estimate relies on low-resolution \herschel\ and JMCT/AzTEC measurements, and therefore may be contaminated (confused) by emission originating from neighboring (unrelated) sources.

\section{Determination of black hole mass and accretion rate, and the past evolution of the SMBH}
\label{SM_sec_M_LLedd}

Using the best-fit values for $\Lop$ and $\fwhb$ (see \S\ref{SM_sec_analysis} above), and applying commonly-used virial estimators of $\mbh$ \cite{Netzer2007_MBH,TrakhtNetzer2012_Mg2}, we obtain $\mbh=6.91\times10^{9}\,\Msun$.
Alternative calibrations of such virial mass estimators do not alter our findings significantly.
For example, the calibration obtained by a recent reverberation mapping study \cite{Bentz2013_lowL_RL} implies $\mbh=5.68\times10^{9}\,\Msun$; that is, smaller than our fiducial measurement by less than 0.1 dex.
We use the simulations described above (\S\ref{SM_sec_analysis}) to estimate the measurement uncertainties on $\mbh$.
The best-fitting $\Lop$ and $\fwhb$ for each simulated spectrum were combined to provide a set of artificial estimates of $\mbh$.
The cumulative distribution function of these $\mbh$ estimates is presented in the lower-right panel of Fig.~\ref{fig:CID_947_sim}.
For the $S/N=10$ simulations we find that the 16\% and 84\% quantiles are at $5.73\times10^{9}$ and $7.66\times10^{9}\,\Msol$, respectively, resulting in $1\sigma$-equivalent uncertainties on $\mbh$ of $-1.18\times10^{9}$ and $+0.75\times10^{9}\,\Msol$.
We further find that 95\% (99\%) of these simulations resulted in $\mbh>5.09\times10^{9}\,\Msol$ ($3.61\times10^{9}\,\Msol$). 
The corresponding values for the $S/N=5$ simulations are $\mbh>3.82\times10^{9}$ and $2.96\times10^{9}\,\Msol$, respectively. 
Virial (or ``single-epoch''), \Hbeta-based estimates of $\mbh$ are also known to be prone to systematic uncertainties, of up to $\sim0.4$ dex, due to the reliance on the empirical $\RBLR-\Lop$ relation, and the overall normalization of the mass estimators.
We stress however that our analysis of \mysobj\ should, in principle, suffer \emph{less} from systematics, compared to similar studies of more luminous sources at $z\sim3.5$, as its luminosity of $\Lop=3.4\times10^{45}\,\ergs$ lies within the range covered directly by reverberation mapping experiments \cite{Kaspi2005,Bentz2009_RL_host,Bentz2013_lowL_RL}.
The leading systematic uncertainty in the present case is therefore associated with the assumption of a typical ``geometrical factor'' (commonly referred to as $f_{\rm BLR}$), which is of order 0.1 dex \cite{Grier2013_sigs_PGs,Woo2013_RM_Msig}.

Another source of concern is the possibility that the BLR is observed at a high inclination angle, as suggested by the presence of the broad \emph{absorption} features (i.e., BAL features) in the rest-frame UV part of the spectrum of \mysobj\ (see \S\ref{SM_sec_BAL} below). 
One may suspect that in such a case, the measured $\fwhb$ would systematically over-estimate the typical velocity dispersion in the BLR, and thus lead to an overestimated $\mbh$.
We have investigated this issue by comparing the distributions of $\fwmg$ in large samples of non-BAL QSOs, and those with a BAL feature in the \CIV\ line (as in \mysobj), among sources at $1.6<z<1.9$, drawn from a large catalog based on the Sloan Digital Sky Survey \cite{Shen_dr7_cat_2011}.
Although in the current work we used \Hbeta, and not the \MgII\ line as our virial estimator, the widths of these two lines have been shown to be closely correlated, and they are thought to originate from a similar region within the BLR (e.g., \cite{Shen_dr7_cat_2011,TrakhtNetzer2012_Mg2}).
The distributions of $\fwmg$ for BAL and non-BAL QSOs (1723 and 15370 objects, respectively) are very similar in shape, with the median FWHM value for BAL QSOs being only slightly higher, by merely $230\,\kms$. 
Moreover, there is no excess of BAL QSOs with line widths comparable to what we estimate for \mysobj\ (i.e., $\gtsim10,000\,\kms$).

Several phenomenological studies have raised the possibility that, for \emph{some} luminous sources with particularly broad \Hbeta\ lines ($\fwhb\gtsim4000\,\kms$), the line profiles might include a significant contribution from a non-virialized ``very broad component'', which should not be taken into account when estimating $\mbh$ (e.g., \cite{Marziani2013_MgII}). 
Simply adopting the empirically derived (and perhaps luminosity-dependent) corrections suggested in such studies \cite{Marziani2009,Marziani2013_MgII}, our estimate for $\mbh$ should be scaled down by about 0.2-0.25 dex. 
In the context of our main finding, of an extremely high BH-to-stellar mass ratio for \mysobj, this would mean $\mmsmall\simeq1/15 - 1/12$ - still a very high value (see Fig.~\ref{fig:MM_vs_z}).
We however stress that the commonly advocated approach to singling out such peculiar objects is based on the unambiguous identification of two components in the broad \Hbeta\ emission line (the ``core'' and the ``very broad component''), as well as some line asymmetry.
Since the \Hbeta\ profile in \mysobj\ does not show such a complicated structure, the aforementioned empirical corrections should not be applied.

Finally, the alternative, ``one sided'' estimate of the line width (see \S\ref{SM_sec_analysis} above) would translate to a decrease of about 0.18 dex in any ``virial'' estimate of $\mbh$. 
As explained in \S\ref{SM_sec_analysis}, and demonstrated in our simulations, this is not a robust estimate of the line width. 
We note, however, that such a decrease in $\mbh$ would have a similar effect on our main result as the one discussed above, namely providing $\mmsmall\simeq1/12$.

We conclude that the SMBH powering the AGN in \mysobj\ has a mass of $\mbh > 3.6\times10^9\,\Msun$, at the 99\% confidence level, and our best estimate is $\mbh=6.9\times10^{9}\,\Msun$.

The rest-frame optical spectrum was used to derive yet another estimate of $\Lbol$, by applying a bolometric correction (i.e., $\fbolopt\equiv\Lbol/\Lop$).
This approach is consistent with many previous studies of un-obscured AGN, at all redshifts.
We used the luminosity-dependent prescription described in \cite{TrakhtNetzer2012_Mg2}, which in turn relies on the $B$-band bolometric corrections presented in \cite{Marconi2004}, translated to 5100 \AA\ assuming a UV-optical SED with $f_{\nu}\propto \nu^{-1/2}$ \cite{VandenBerk2001}.
For \mysobj, this results in $L_{\rm bol,\, opt}=2.11\times10^{46}\,\ergs$.
This value is highly consistent with the \xmm-based estimate of $L_{\rm bol,\, SED}$ (ref.\ \cite{Lusso2011_Lbol}; within 0.06 dex), but significantly higher than the \chandra-based value (ref.\ \cite{Elvis2012_SED}; by a factor of almost 2).

The derived values of $\Lbol$ and $\mbh$ were combined to provide estimates of the normalized accretion rate, in terms of the Eddington luminosity, $\lledd \equiv \Lbol / \left(1.5\times10^{38} \, \mbh/\Msun\right)$ (this definition of the ``Eddington ratio'' assumes a Solar gas-phase metalicity).
We obtain $\lledd=0.021$ for the $\Lop$-based estimate of $\Lbol$, or $0.019$ and $0.011$ for the \xmm\ and \chandra-based estimates of $L_{\rm bol,\, SED}$, respectively.
Given the fact that the estimates of $\Lbol$ were obtained using very different approaches, and the systematic uncertainties associated with the estimation of $\Lbol$ (e.g., the scatter in $\fbolopt$), we consider these estimates of $\lledd$ to be in excellent qualitative agreement: 
the SMBH in \mysobj\ is accreting at a rate of at most $\lledd\simeq0.02$.

By combining the estimated $\lledd$ and a standard radiative efficiency of $\eta=0.1$ \cite{Marconi2004}, we obtain an $e$-folding timescale for the growth of the SMBH, following the expression 
$\tau_{\rm BH} = 4 \times 10^8 ~~ \frac {\eta /(1- \eta) }{\lledd} ~ {\rm yr}$.
The resulting timescales are about 2.1 and 4 Gyr, for the higher ($\Lop$-based) and lower (\chandra-based) estimates of $\lledd$, respectively.
In any case, these timescales are longer than the age of the Universe at $z=3.328$, of about 1.88 Gyr, and than the elapsed time since the earliest seed black holes likely formed,  1.7 Gyr ($z\sim20$).
This very long timescale thus suggests that the SMBH in \mysobj\ had to experience an earlier epoch of much faster growth (i.e., higher accretion rate), and/or that it had to originate from the most massive type of seeds.
In particular, extrapolating the growth history ``backwards'' to $z=10$, assuming constant accretion rate and radiative efficiency of $\lledd=0.02$ and $\eta=0.1$ (respectively), results in $\mbh\left(z=10\right)\simeq 3.7\times10^{9}\,\Msun$. 
Even with the lowest efficiency within the framework of a geometrically-thin, optically thick accretion disk, $\eta=0.038$ (maximally retrograde spinning BH; e.g., \cite{Netzer2013_book}), the implied mass is still about $10^{9}\,\Msun$. 
The most extreme BH seed production mechanisms rely on different ``direct collapse'' scenarios, but generally provide very few seeds as massive as $\mseed\sim10^6\,\Msun$ (see, e.g., \cite{Bonoli2013} and reviews in \cite{Volonteri2010_rev,Natarajan2011_seeds_rev,Volonteri2012_Science_rev}). 
Some very recent studies speculate that \emph{some} BH seed masses may be yet higher, perhaps up to $\sim10^{8}\,\Msun$, but not before $z=10$ \cite{Mayer2015_direct}.
Considering the accretion rate onto the SMBH, some recent models highlight the possibility of yet more efficient accretion, as the disk becomes ``slim'' and surpasses the simplified (spherical) Eddington limit, perhaps reaching $\dot{M}/M_{\rm Edd}\sim3$ \cite{Madau2014_supEdd}.
We note however that such extreme models for BH seed production and accretion may not necessarily be required to explain objects like \mysobj.
The implied $\mbh$ of \mysobj\ at $z\sim5$, of about $3\times10^{9}\,\Msun$, is consistent with that observed in the population of high-luminosity quasars at that epoch \cite{Trakhtenbrot2011,DeRosa2011}. 
Such sources, however, have much higher accretion rates, typically $\lledd\gtsim0.5$, and can emerge from standard accretion (at relatively high duty cycles and low radiative efficiency), from a broad range of BH seed masses, including those of stellar remnants (i.e., $\mseed\ltsim10^3\,\Msun$). 
We stress that in any case, \mysobj\ had to have higher-than-observed accretion rate \emph{sometime} in the past, to account for its high mass.
As we show in \S\ref{SM_sec_BAL} below, we have evidence that this epoch of high accretion rate took place relatively shortly before the observed epoch.

\section{Rest-frame UV spectrum and BAL features}
\label{SM_sec_BAL}

An optical spectrum of CID-947 was obtained as part of the zCOSMOS survey (ref.\ \cite{Lilly2007_zCOSMOS}; zCOSMOS-ID 823936), and we present it in Fig.~\ref{fig:CID_947_zCOSMOS}.
It clearly shows very broad and deep absorption troughs blue-ward of the \SilIVuv\ and \CIV\ lines, identifying \mysobj\ as a Broad Absorption Line Quasar, a sub-population that comprises about 10-20\% of luminous, un-obscured AGN (BALQSOs; see, e.g.,\cite{Weymann1981_BAL_rev,Crenshaw2003_ARAA,Gibson2009_BAL_cat}, and references therein). Moreover, the absorption feature blue-ward of the \AlIII\ line suggests that \mysobj\ may belong to the yet rarer sub-class of low-ionization BAL QSOs (i.e., it is a ``LoBAL'').
For both $\civ$ and $\Siliv$ we estimate a maximum outflow velocity of $\left| v_{\rm max} \right|\simeq12,000\,\kms$, with the $\civ$ trough probably slightly broader.
While this value of $v_{\rm max}$ is not uncommon among BALQSOs \cite{Gibson2009_BAL_cat}, it \emph{is} an outlier in the $v_{\rm max} - \lledd$ plane:
virtually all known BALQSOs with comparable $v_{\rm max}$ have much higher accretion rates, typically $\lledd>0.1$ \cite{Ganguly2007}.
A simple model for the launching of such high-velocity outflows \cite{Hamann1998,Misawa2007} yields a value for the maximum outflow velocity  
$v_{\rm max}\simeq 9300 \left(\frac{R_{\rm abs, 0.1}}{M_{8}}\right)^{-1/2} \left(1.5\frac{f_{0.1}}{N_{22}}\lledd -  0.1 \right)^{1/2} \,\, \kms$,
where $R_{\rm abs, 0.1}$ is the distance of the absorber from the continuum source, scaled to 0.1 pc; 
$M_{8}$ is the SMBH mass, scaled to $10^8\,\Msol$;
$f_{0.1}$ is the fraction of continuum photons absorbed (or scattered) by the outflowing gas, scaled to 10\%;
and $N_{22}$ is the absorber column density, scaled to $10^{22}\,\cmii$.
Assuming the observed value for $\mbh$, and also $N_{\rm H}=10^{22}\,\cmii$, $R_{\rm abs, 0.1}=1$ and $f_{0.1}=1$, this expression implies that a wind with $v_{\rm max}\simeq12,000\,\kms$ should have been launched by a SMBH accreting at $\lledd\gtsim 0.1$, and probably at rates as high as $\lledd\simeq 0.7$ (see \cite{Misawa2007} for discussion of viable ranges on all parameters).
This is significantly higher, by at least an order of magnitude, than the observed value of $\lledd$.
We also note that, since within the framework of this model $\lledd\propto R_{\rm abs, 0.1}$, the implied accretion rate can easily reach $\lledd\simeq1$ if the outflow has reached $\sim1$ pc.

The high-velocity outflow was launched at a time $\Delta t \sim R_{\rm abs}/v_{\rm max}$ prior to the observed epoch. 
Even for a conservative assumption of $R_{\rm abs}\gtsim 1\,{\rm kpc}$ (e.g., ref.\ \cite{Capellupo2014_BAL} and references therein), the implied age of the outflow is about $10^5$ years. 
Any alternative, more realistic assumption regarding $R_{\rm abs}$ would imply an even shorter timescale.

Thus, the relatively high terminal velocity of the observed outflow lends further support to the scenario in which \mysobj\ was accreting at much higher rates, probably as recently as $\sim10^5$ years before the observed epoch.

\section{Subsequent evolution of $\mstar$}
\label{SM_sec_mstar_evo}

We estimated the ``final'' stellar mass of \mysobj\ ($\mgal\left(z=0\right)$) in several ways, all of which rely on the observed stellar mass ($5.6\times10^{10}\,\Msol$) and star formation rate ($\sim400\,\mpyr$).
Our calculations assume that we are observing the host galaxy of \mysobj\ near its peak of star forming activity, and that the SFR can only decline with time.

First, we assume an exponential decline in SFR, with typical $e$-folding timescales in the range of $\tau=1-2$ Gyr.
These short timescales are supported by several observational studies which constrain the ages of the stellar populations in massive, low-redshift galaxies \cite{Thomas2005}, and basically implies that the final mass of the galaxy can be approximated by 
$\mgal\left(z=0\right) \simeq \mgal \left(z=3.328\right) + {\rm SFR} \times \tau$.
These ``integrated'' masses should be scaled down, by factors of about 1.6, to account for the fact that some of the mass is returned back to the galaxy's gas (e.g., through stellar winds, Supernovae explosions etc.).
Using the observed $\mgal$ and SFR for \mysobj\ we obtain final masses of $\mgal\left(z=0\right) \simeq \left(2.8-5.3\right) \times10^{11}\,\Msol$.

An alternative calculation relies on the scenario in which star-forming galaxies evolve on the ``main sequence'' at all epochs, until they quench, and that the mass functions of such objects, at all epochs, can be linked via the so-called ``continuity approach'' \cite{Peng2010_Qing}.
In particular, we assume that for star forming galaxies the specific star formation rate, sSFR, evolves as ${\rm sSFR}\sim1/t$, and that the probability of quenching depends predominantly on $\mstar$.
The different parameters in the calculations were derived by applying the continuity approach to the observed (evolving) stellar mass functions (see ref. \cite{Peng2010_Qing} for details).
These calculations predict that a galaxy with properties like those of \mysobj\ will evolve to reach $7.2\times10^{11}\,\Msol$ at $z=0$, if it \emph{never} quenches. 
Alternatively, it can reach $2.2\times10^{11}\,\Msol$ if it quenches at $z_{\rm quench}=2.51$, when the (mass-dependent) probability of quenching reaches 75\%.
Allowing for some additional mass growth through galaxy-galaxy mergers would increase the latter estimate of final mass by a factor of about 2.6, to $5.75\times10^{11}\,\Msol$.
The increase due to mergers for the former mass estimate is minute, since the probability of experiencing a similar-mass merger decreases with increasing mass.
In any case, these calculations show that the stellar population in \mysobj\ should grow by about an order of magnitude between $z\simeq3.3$ and $z\sim0$.

We note that the high SFR we measure in \mysobj\ implies a molecular gas mass of at least $M\left({\rm H}_2\right)\simeq2\times10^{10}\,\Msun$, and perhaps as high as $2\times10^{11}\,\Msun$ \cite{Carilli_ARAA_2013,Sargent2014}. 
This suggests that the host galaxy can experience significant growth by consuming this reservoir of cold gas, without any additional gas accretion from its surroundings (i.e., the IGM), nor from mergers.

\clearpage

\begin{figure*}[h]
\centering
\includegraphics[angle=-90,width=0.95\linewidth]{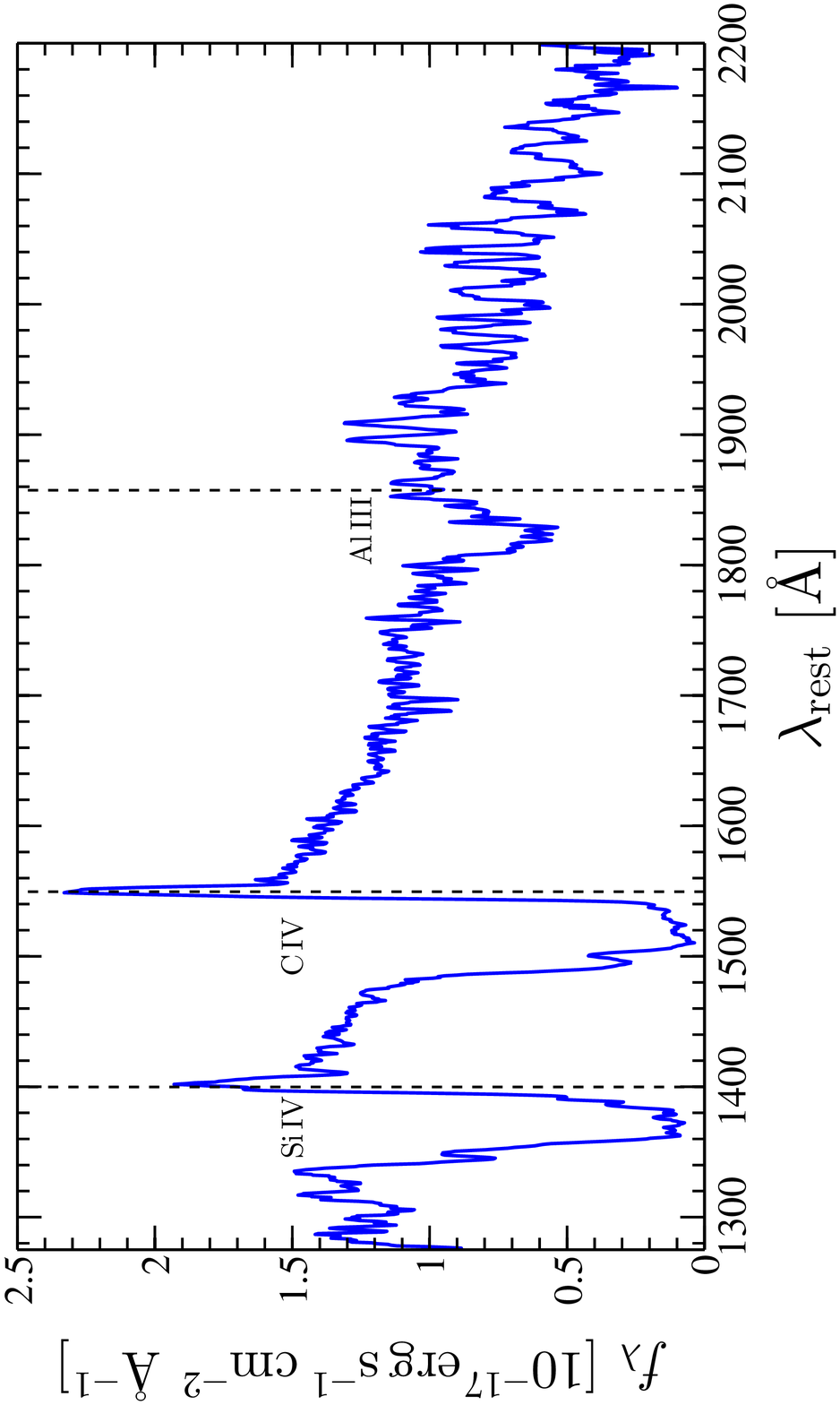}
\caption{
Optical spectrum of \mysobj, probing the rest-frame UV regime, obtained as part of the zCOSMOS survey \cite{Lilly2007_zCOSMOS}.
Dashed vertical lines mark the expected center wavelength of the \SilIVuv, \CIV, and \AlIII\ lines. 
Blue-shifted, broad absorption features are clearly detected next to each of these three lines, identifying \mysobj\ as a low-ionization broad absorption line QSO (or ``LoBAL QSO'').
The broad absorption troughs blue-ward of \Siliv\ and \civ\ reach $v_{\rm max}\simeq12000\,\kms$.
The red part of the spectrum ($ \lambda _{\rm rest} \le 1900$ \AA), including the \CIII\ line, is known to be affected by (instrumental) fringing.
}
\label{fig:CID_947_zCOSMOS}
\end{figure*}

\clearpage

\begin{table}
\centering
\caption{UV-to-IR Spectral Energy Distribution}\label{tab:sed}
\begin{center}
\begin{tabular}{llcccc}
\hline
\hline
 Telescope/ & band & $\lambda_{\rm obs}$ & $m_{\rm AB}\pm\Delta$        & $\lambda_{\rm rest}$  & $\log\left(\lambda L_{\lambda}\right)\pm\Delta$ \\
 Instrument & ~~ & ~~~~                  & ~~~~                         & $\left[\mu {\rm m}\right]$   & $\left[\ergs\right]$ \\
\hline
Subaru/    & $V$	& $5449$ \AA	& $20.900\pm0.007$	& $0.1259$	& $45.937\pm0.003$	\\
SuprimeCam & $r$	& $6232$ \AA	& $20.803\pm0.006$	& $0.1440$	& $45.918\pm0.002$	\\
~~~~~~~~~~ & $i^{+}$	& $7621$ \AA	& $20.394\pm0.004$	& $0.1761$	& $45.994\pm0.002$	\\
~~~~~~~~~~ & $z^{++}$	& $8801$ \AA	& $20.588\pm0.002$	& $0.2033$	& $45.854\pm0.001$	\\
VISTA/     & $Y$	& $1.020$ \mic	& $20.639\pm0.003$	& $0.2357$	& $45.770\pm0.001$	\\
VIRCAM     & $J$	& $1.250$ \mic	& $20.333\pm0.003$	& $0.2888$	& $45.803\pm0.001$	\\
~~~~~~~~~~ & $H$	& $1.650$ \mic	& $20.290\pm0.004$	& $0.3813$	& $45.700\pm0.001$	\\
~~~~~~~~~~ & $K_{\rm s}$	& $2.154$ \mic	& $20.029\pm0.005$	& $0.4976$	& $45.689\pm0.002$	\\
\spitzer/  & ch1   & $3.526$ \mic	& $19.734\pm0.011$	& $0.8147$	& $45.593\pm0.004$	\\
IRAC~~~~~~ & ch2   & $4.461$ \mic	& $19.589\pm0.009$	& $1.0307$	& $45.549\pm0.003$	\\
~~~~~~~~~~ & ch3   & $5.677$ \mic	& $19.349\pm0.073$	& $1.3117$	& $45.540\pm0.027$	\\
~~~~~~~~~~ & ch4   & $7.704$ \mic	& $18.765\pm0.075$	& $1.7800$	& $45.641\pm0.028$	\\
MIPS~~~~~~ & 24${\mu}$   & $23.68$ \mic	& $16.868\pm0.026$	& $5.4704$	& $45.912\pm0.010$	\\
\hline
\end{tabular}
\end{center}
\end{table}

\end{document}